\documentclass[12pt]{article}
\usepackage{amsmath,amssymb,amsfonts, amsbsy, epsfig, subfig}
\usepackage[usenames]{color}
\usepackage{rotating}
\usepackage{boxedminipage}

\usepackage{rotating}
\newcommand{\ignore}[1]{}{}

\newtheorem{theorem}{Theorem}[section]
\newtheorem{corollary}{Corollary}[section]
\newtheorem{proposition}{Proposition}[section]
\newtheorem{lemma}{Lemma}[section]
\newtheorem{example}{Example}[section]

\newtheorem{definition}{Definition}[section]

\newcommand{\beq}{\begin{equation}}
\newcommand{\eeq}{\end{equation}}
\newcommand{\beas}{\begin{eqnarray*}}
\newcommand{\eeas}{\end{eqnarray*}}
\newcommand{\bea}{\begin{eqnarray}}
\newcommand{\eea}{\end{eqnarray}}
\newcommand{\bei}{\begin{itemize}}
\newcommand{\eei}{\end{itemize}}
\newcommand{\ben}{\begin{enumerate}}
\newcommand{\een}{\end{enumerate}}
\newcommand{\bet}{\begin{theorem}}
\newcommand{\eet}{\end{theorem}}
\newcommand{\bel}{\begin{lemma}}
\newcommand{\eel}{\end{lemma}}
\newcommand{\bep}{\begin{proposition}}
\newcommand{\eep}{\end{proposition}}
\newcommand{\bed}{\begin{definition}}
\newcommand{\eed}{\end{definition}}
\newcommand{\bec}{\begin{corollary}}
\newcommand{\eec}{\end{corollary}}
\newcommand{\bex}{\begin{example}}
\newcommand{\eex}{\end{example}}
\newcommand{\qed}{\quad\hbox{\vrule width 4pt height 6pt depth 1.5pt}}

\newcommand{\ep}{\epsilon}

\newcommand{\argmax}{\mathop{\rm arg\max}}

\def\0{\boldsymbol{0}}
\def\F{\boldsymbol{F}}
\def\E{\boldsymbol{E}}

\def\R{\boldsymbol{R}}

\def\A{\boldsymbol{A}}

\def\X{\boldsymbol{X}}
\def\I{\boldsymbol{I}}

\def\m{\boldsymbol{\mu}}
\def\0{\boldsymbol{0}}

\def\va{\boldsymbol{\varepsilon}}
\def\X{\boldsymbol{X}}

\def\m{\boldsymbol{\mu}}

\def\pr{\textsf{P}} 
\def\ep{\textsf{E}} 
\def\Var{\textsf{Var}} 

\begin{document}

\title{Incorporation of Sparsity Information in Large-scale Multiple Two-sample $t$ Tests
}
\author{Weidong Liu\footnote{Department of Mathematics, Institute of Natural Sciences and MOE-LSC, Shanghai Jiao Tong University. Research supported by NSFC, Grants No.11201298, No.11322107 and No.11431006, Program for New Century Excellent Talents in
University,  Shanghai Pujiang Program, 973 Program and  a grant from Australian Research Council. Email: weidongl@sjtu.edu.cn. }}

\maketitle

\begin{abstract}
Large-scale multiple two-sample {\em Student}'s $t$ testing problems often arise from the statistical analysis of  scientific data. To detect components with different values between two mean vectors, a well-known procedure is to apply the  Benjamini and Hochberg (B-H)  method and two-sample {\em Student}'s $t$ statistics to control the false discovery rate (FDR). In many applications, mean vectors  are  expected to be sparse or asymptotically sparse.
When dealing with such type of data, {\em can we gain more power than the standard procedure such as the B-H method with Student's $t$ statistics while keeping the FDR under control?} The answer is positive. By exploiting the possible sparsity  information in  mean vectors,
we present an uncorrelated screening-based (US) FDR control procedure, which is shown to be more powerful than the B-H method.  The US testing procedure depends on a novel construction of screening statistics, which are asymptotically uncorrelated with  two-sample {\em Student}'s $t$  statistics. The US testing procedure  is different from some existing {\em testing following screening} methods (Reiner, et al., 2007; Yekutieli, 2008) in which  independence between screening and testing  is crucial to  control the FDR, while the independence  often requires additional data or splitting of samples. An inappropriate splitting of samples may result in a loss  rather than an improvement of statistical power.  Instead, the uncorrelated screening US is based on the original data and does not need to split the samples. Theoretical results show that the US testing procedure controls the desired FDR asymptotically.
 Numerical studies are conducted and indicate that the proposed procedure works quite well.
\end{abstract}


\section{Introduction}

Modern statistical analysis of high-dimensional data often involves multiple two-sample  hypothesis tests
 \begin{eqnarray*}
 H_{i0}:~\mu_{i,1}=\mu_{i,2}\quad\mbox{versus}\quad H_{i1}:~\mu_{i,1}\neq \mu_{i,2},\quad 1\leq i\leq m,
 \end{eqnarray*}
 where $\m_{1}=(\mu_{1,1},\ldots,\mu_{m,1})$ and $\m_{2}=(\mu_{1,2},\ldots,\mu_{m,2})$ are two population mean vectors and $m$ usually can  be tens of thousands. Ever since the seminal work of Benjamini and Hochberg (1995), the false discovery rate (FDR) control is becoming more and more desirable in large-scale multiple testing problems. The concept of FDR control not only provides an easily accessible measure on the overall type I error but also allows higher  statistical power than the conservative family-wise error rate control.
Let $p_{1},\ldots,p_{m}$ be  p-values calculated from two-sample {\em Student}'s statistics for $H_{10},\ldots,H_{m0}$, respectively. The well known Benjamini and Hochberg (B-H) method rejects $H_{j0}$ if $p_{j}\leq p_{(\hat{k})}$, where
\begin{eqnarray*}
\hat{k}=\max\{0\leq k\leq m:~p_{(k)}\leq \alpha k/m\}
\end{eqnarray*}
and $p_{(1)}<\ldots<p_{(m)}$ are the order p-values. Benjamini and Hochberg (1995) prove that their procedure controls the FDR at level $\alpha$ if $p_{1},\ldots,p_{m}$ are independent. After their seminal work, there are a huge amount of literature on the FDR control under various settings; see
Benjamini and Yekutieli (2001), Efron (2004,2007), Storey (2003), Storey, et al. (2004), Ferreira and Zwinderman (2006), Wu (2008), Sun and Cai (2009), Cai, et al. (2011) and so on.

In many applications, the mean vectors $\m_{1}$ and $\m_{2}$ are expected to be sparse or asymptotically sparse. For example, in genetics, a quantitative trait could be controlled by a few major genes and many polygenes, and it is typically assumed that the polygenes
have vanishingly small effects. In  genome-wide association studies (GWAS),  by marginal regressions,
Fan, et al. (2012) convert GWAS into large-scale multiple testing $H_{i0}:$ $\mu_{i}=0$, $1\leq i\leq m$, for a mean vector $(\mu_{1},\ldots,\mu_{m})$ of $m$-dimensional normal random vector,  where $\mu_{i}$ denotes the correlation coefficient between the $i$-th SNPs and a response such as genetic traits or disease status. It is reasonable to assume that only a few SNPs contribute to the response so that $(\mu_{1},\ldots,\mu_{m})$
is expected to be asymptotically sparse. In the estimation of high-dimensional mean vectors and the context of signal detections,  mean vectors are also often assumed to be sparse; see Abramovich, et al. (2006), Cai and Jeng (2011) and Donoho and Jin (2004).

 When  $\m_{1}$ and $\m_{2}$ are (asymptotically) sparse,
 {\em can we gain more power than standard procedures such as the B-H method with Student's $t$ statistics while keeping the FDR under control?} The answer is trivially positive if the union support $\mathcal{S}_{1}\cup\mathcal{S}_{2}$ of $\m_{1}$ and $\m_{2}$ is known and small, where $\mathcal{S}_{j}=\{i: \mu_{i,j}\neq 0\}$, $j=1,2$. Actually, the support of $\m_{1}-\m_{2}$ is contained in $\mathcal{S}_{1}\cup\mathcal{S}_{2}$. Applying the B-H method to those components with indices in $\mathcal{S}_{1}\cup\mathcal{S}_{2}$ will significantly improve the statistical power. The union support $\mathcal{S}_{1}\cup\mathcal{S}_{2}$ is of course unknown and can even be as large as $\{1,2,\ldots,m\}$ if $\m_{1}$ and $\m_{2}$ are asymptotically sparse.
  One may  screen the mean vectors to obtain an estimate for the union support in the first stage and test the set of identified hypotheses  while controlling the FDR in the second stage. This is known as {\em testing following screening} method which has been used in other multiple testing problems; see  Zehetmayer, et al. (2005), Reiner, et al. (2007) and Yekutieli (2008). For such method, independence between screening in the first stage and  hypothesis testing  in the second stage is crucial to control the FDR. If independence is absent, by a simulation study, Reiner, et al. (2007) show that  when hypotheses are screened by
1-way ANOVA $F$ tests, the B-H procedure is unable to control the FDR in the second step as  p-values no longer remain Uniform $(0,1)$.
In Section 4, we will further state some simulation results and show that it is impossible to control the FDR with the B-H method and some seemingly natural screening statistics. The independence between screening and hypothesis testing often requires additional data or splitting of samples. In the latter approach, it is difficult to determine the reasonable fractions of  samples in two stages and the result may be unstable in real data applications. Moreover, a simulation study in Section 4 indicates that an inappropriate splitting of samples may result in a loss of statistical power.

In this paper, we present an uncorrelated screening-based (US) testing procedure for the FDR control,  by a novel construction of screening statistics which are asymptotically uncorrelated with two-sample {\em Student}'s $t$  statistics. Instead of the independence assumption between screening and testing, we show that in the US  procedure, an asymptotic zero correlation  is sufficient for the FDR control. The US procedure does not require any other samples or splitting of samples.
It is demonstrated that the proposed US  procedure is more powerful than the classical B-H method while keeping
the FDR controlled at the desired level.
Particularly,  we prove that the range of signal sizes,  in which the power of  the US  procedure converges to one, is wider than that of the B-H method, by exploiting the possible sparsity  information in  mean vectors. The asymptotic sparsity assumption for the power results in Section 3.2 is quite weak.
 It allows $m^{\gamma}$, $0<\gamma<1$, components of mean vectors that can be arbitrarily large. The remaining components can be of the order of $\sqrt{\log m/n}$, which may still be moderately large for ultra-high dimensional settings, for example, $\log m\geq cn$ for some $c>0$. On the unfavorable case that
$\m_{1}$ and $\m_{2}$ are non-sparse at all,
 the US procedure will still be at least as powerful as the B-H method. That is, the US procedure does not really require the sparsity assumption on $\m_{1}$ and $\m_{2}$. But if they share asymptotic sparsity, then US procedure can incorporate this information and improves statistical power.

 We shall note that
the exact null distributions of two-sample {\em Student}'s $t$ statistics are typically unknown. The US procedure  does not require the true null distributions. Instead,
our results show that it is robust to the asymptotic null distributions under some moment conditions. Moreover, our results  allow  $K$-dependence between the components of populations.

The remainder of this paper is organized as follows. In Section 2, we introduce the US testing procedure. Section 3 gives  theoretical results on the FDR and FDP control. Theoretical comparisons between the US procedure and the B-H method are also given. The simulation study is presented in Section 4 and a discussion on  several possible extensions is given in Section 5. The proofs of main results are postponed to Section 6.
Throughout, we let $C$ and $C_{(\cdot)}$ denote positive constants which may be different in each place. For two  sequences of real numbers
$\{a_{m}\}$ and $\{b_{m}\}$, write $a_{m} = O(b_{m})$ if there
exists a constant $C$ such that $|a_{m}| \leq C|b_{m}|$ holds for all
sufficiently large $m$, and write $a_{m} = o(b_{m})$ if
$\lim_{m\rightarrow\infty}a_{m}/b_{m} = 0$. For a set $\A\subset\{1,2,\ldots,m\}$, $|\A|$ denotes its cardinality.

\section{Uncorrelated screening-based  FDR control procedure}

In this section, we introduce the US testing procedure.
Let $\mathcal{X}_{1}:=\{\X_{k,1},1\leq k\leq n_{1}\}$ and $\mathcal{X}_{2}:=\{\X_{k,2},1\leq k\leq n_{2}\}$ be i.i.d. random samples from $\X_{1}$ and $\X_{2}$, respectively, where $\m_{1}=\ep\X_{1}$ and $\m_{2}=\ep \X_{2}$. Assume that $\mathcal{X}_{1}$ and $\mathcal{X}_{2}$ are independent. Set
\begin{eqnarray*}
\X_{k,1}=(X_{k,1,1},\ldots,X_{k,m,1})\quad\mbox{and\quad}\X_{k,2}=(X_{k,1,2},\ldots,X_{k,m,2}).
\end{eqnarray*}
Let $\X_{j}=(X_{1,j},\ldots,X_{m,j})$, $j=1,2$. The variances $\sigma^{2}_{i,1}=\Var(X_{i,1})$ and $\sigma^{2}_{i,2}=\Var(X_{i,2})$, $1\leq i\leq m$.

{\em Case I, equal variances $\sigma^{2}_{i,1}=\sigma^{2}_{i,2}$, $1\leq i\leq m$.} We define two-sample {\em Student}'s $t$ statistic for $H_{i0}$ by
\begin{eqnarray*}
T_{i}=\sqrt{\frac{n_{1}n_{2}}{(n_{1}+n_{2})\hat{\sigma}^{2}_{i,pool}}}(\bar{X}_{i,1}-\bar{X}_{i,2}),
\end{eqnarray*}
where
\begin{eqnarray*}
\bar{X}_{i,j}=\frac{1}{n_{j}}\sum_{k=1}^{n_{j}}X_{k,i,j}\mbox{\quad and\quad}\hat{\sigma}^{2}_{i,pool}=\frac{1}{n_{1}+n_{2}-2}\sum_{j=1}^{2}\sum_{k=1}^{n_{j}}(X_{k,i,j}-\bar{X}_{i,j})^{2}
\end{eqnarray*}
for $j=1,2$.
The key step in the US testing procedure is the construction of an  uncorrelated screening statistic which can  screen out nonzero components.
In equal variances case, the US procedure  uses
\begin{eqnarray*}
S_{i}=\sqrt{\frac{n^{2}_{1}}{(n_{1}+n_{2})\hat{\sigma}^{2}_{i,pool}}}(\bar{X}_{i,1}+\frac{n_{2}}{n_{1}}\bar{X}_{i,2})
 \end{eqnarray*}
 as a screening statistic.

 {\em Case II,  variances $\sigma^{2}_{i,1}$ and $\sigma^{2}_{i,2}$ are not necessary equal.} In this case, we define two-sample {\em Student}'s $t$ statistic
 \begin{eqnarray*}
T_{i}=\frac{\bar{X}_{i,1}-\bar{X}_{i,2}}{\sqrt{\hat{\sigma}^{2}_{i,1}/n_{1}+\hat{\sigma}^{2}_{i,2}/n_{2}}},
\mbox{\quad where\quad}
\hat{\sigma}^{2}_{i,j}=\frac{1}{n_{j}-1}\sum_{k=1}^{n_{j}}(X_{k,i,j}-\bar{X}_{i,j})^{2}
\end{eqnarray*}
for $j=1,2$.
The US procedure  uses
\begin{eqnarray*}
S_{i}=\sqrt{\frac{n_{1}}{\hat{\sigma}^{2}_{i,1}(1+\frac{n_{2}\hat{\sigma}^{2}_{i,1}}{n_{1}\hat{\sigma}^{2}_{i,2}})}}(\bar{X}_{i,1}+\frac{n_{2}\hat{\sigma}^{2}_{i,1}}{n_{1}\hat{\sigma}^{2}_{i,2}}\bar{X}_{i,2})
 \end{eqnarray*}
 as a screening statistic.

 The construction of screening statistic is quite straightforward, but the idea can be extended to many other two-sample testing problems.
 Note that $S_{i}$ is asymptotically equivalent to $$S_{0i}:=\sqrt{\frac{n_{1}}{\sigma^{2}_{i,1}(1+n_{2}\sigma^{2}_{i,1}/(n_{1}\sigma^{2}_{i,2}))}}
 (\bar{X}_{i,1}+\frac{n_{2}\sigma^{2}_{i,1}}{n_{1}\sigma^{2}_{i,2}}\bar{X}_{i,2}),$$ which is uncorrelated with $\bar{X}_{i,1}-\bar{X}_{i,2}$.
 Note that $$\ep S_{0i}=\sqrt{\frac{n_{1}}{\sigma^{2}_{i,1}(1+n_{2}\sigma^{2}_{i,1}/(n_{1}\sigma^{2}_{i,2}))}}(\mu_{i,1}+\frac{n_{2}\sigma^{2}_{i,1}}{n_{1}\sigma^{2}_{i,2}}\mu_{i,2}).$$
Hence, $S_{i}$ can  filter out zero components while keeping nonzero components.
If the signs of $\mu_{i,1}$ and $\mu_{i,2}$ are opposite, then  $\ep S_{0i}$ can be small.
However, we do not need to care about this case. It will  always be easier for $T_{i}$ to detect components with $\mu_{i,1}\mu_{i,2}<0$ than those with the same signal sizes but  $\mu_{i,1}\mu_{i,2}>0$, because signals in the first case are stronger  than signals in the latter case  in terms of $\mu_{i,1}-\mu_{i,2}$. For the components which haven't been selected by $S_{i}$, a separate multiple testing  will be applied on them.

We use $\Psi(t)$, the {\em Student}'s $t$ distribution with $n_{1}+n_{2}-2$ degrees of freedom,  as an  asymptotic null distribution for $T_{i}$. It is clearly that other distributions such as the normal distribution or bootstrap empirical null distribution can be used. Suppose that we threshold  $|S_{i}|$ at level $\lambda$
and divide $H_{i0}$, $1\leq i\leq m$, into two families $\{H_{i0}:~|S_{i}|\geq \lambda\}$ and $\{H_{i0}:~|S_{i}|< \lambda\}$, where the final choice of $\lambda$ relies on a data-driven method so that it will be a  random variable. To illustrate the idea briefly, we temporarily let $\lambda>0$ be an non-random number. We now apply FDR control procedures to these two families of hypotheses.
Let $\mathcal{B}_{1}=\{i: |S_{i}|\geq \lambda\}$ and $\mathcal{B}_{2}=\mathcal{B}_{1}^{c}$. For $i\in\mathcal{B}_{1}$,
 we reject $H_{i0}$ if $|T_{i}|\geq t_{1}$ for some $t_{1}>0$, and for $i\in\mathcal{B}_{2}$, reject  $H_{i0}$ if $|T_{i}|\geq t_{2}$ for some $t_{2}>0$.
 Define the false discovery proportions for the two families of hypotheses by
 \begin{eqnarray*}
&& FDP_{1,\lambda}(t)=\frac{\sum_{i\in\mathcal{H}_{0}}I\{|S_{i}|\geq \lambda,|T_{i}|\geq t\}}{\max(1,\sum_{i=1}^{m}I\{|S_{i}|\geq \lambda,|T_{i}|\geq t\})},\cr
 &&FDP_{2,\lambda}(t)=\frac{\sum_{i\in\mathcal{H}_{0}}I\{|S_{i}|< \lambda,|T_{i}|\geq t\}}{\max(1,\sum_{i=1}^{m}I\{|S_{i}|< \lambda,|T_{i}|\geq t\})},
 \end{eqnarray*}
 where $I\{\cdot\}$ is an indicator function and $\mathcal{H}_{0}=\{1\leq i\leq m:~\mu_{i,1}=\mu_{i,2}\}$.
 To control the FDR/FDP at level $\alpha$ for these two families, as the B-H method, the ideal choices for $t_{1}$ and $t_{2}$ are
 \begin{eqnarray*}
 \hat{t}^{o}_{1}=\inf\Big{\{}t\geq 0: ~FDP_{1,\lambda}(t)\leq \alpha\Big{\}}\quad\mbox{and\quad}
 \hat{t}^{o}_{2}=\inf\Big{\{}t\geq 0: ~FDP_{2,\lambda}(t)\leq\alpha\Big{\}},
 \end{eqnarray*}
 respectively. It is clearly $\sum_{i\in\mathcal{H}_{0}}I\{|S_{i}|\geq \lambda,|T_{i}|\geq t\}$ and $\sum_{i\in\mathcal{H}_{0}}I\{|S_{i}|< \lambda,|T_{i}|\geq t\}$ are unknown. Since $S_{i}$ is asymptotically uncorrelated with $T_{i}$, we will show that under certain conditions,  the above two terms can be approximated by $\hat{m}_{1,\lambda}^{o}(2-2\Psi(t))$
 and $\hat{m}_{2,\lambda}^{o}(2-2\Psi(t))$, where $$\hat{m}_{1,\lambda}^{o}=\sum_{i\in\mathcal{H}_{0}}I\{|S_{i}|\geq \lambda\}\quad\mbox{and}\quad \hat{m}_{2,\lambda}^{o}=m_{0}-\hat{m}_{1,\lambda}^{o}$$ with $m_{0}=|\mathcal{H}_{0}|$. It is straightforward to bound them by $\hat{m}_{1,\lambda}(2-2\Psi(t))$
 and  $\hat{m}_{2,\lambda}(2-2\Psi(t))$, where
 \begin{eqnarray*}
 \hat{m}_{1,\lambda}=\sum_{i=1}^{m}I\{|S_{i}|\geq \lambda\},\quad \hat{m}_{2,\lambda}=m-\hat{m}_{1,\lambda}.
 \end{eqnarray*}
Using $\hat{m}_{1,\lambda}$ and $\hat{m}_{2,\lambda}$, we introduce the FDR control procedure as follow.

{\em FDR control with US testing.}
Let
 \begin{eqnarray*}
 \hat{t}_{1,\lambda}&=&\inf\Big{\{}t\geq 0: ~\frac{\hat{m}_{1,\lambda}(2-2\Psi(t))}{\max(1,\sum_{i=1}^{m}I\{|S_{i}|\geq \lambda,|T_{i}|\geq t\})}\leq \alpha\Big{\}},\cr
 \hat{t}_{2,\lambda}&=&\inf\Big{\{}t\geq 0: ~\frac{\hat{m}_{2,\lambda}(2-2\Psi(t))}{\max(1,\sum_{i=1}^{m}I\{|S_{i}|<\lambda,|T_{i}|\geq t\})}\leq \alpha\Big{\}}.
 \end{eqnarray*}
We reject those $H_{i0}$ if $i\in\mathcal{R}_{\lambda}$, where
$$\mathcal{R}_{\lambda}=\Big{\{}1\leq i\leq m:~I\{|S_{i}|\geq \lambda,|T_{i}|\geq \hat{t}_{1,\lambda}\}=1\mbox{~or~}I\{|S_{i}|< \lambda,|T_{i}|\geq \hat{t}_{2,\lambda}\}=1\Big{\}}.$$
Note that if $\mu_{i,1}=\mu_{i,2}=0$, then $\pr(|S_{i}|\geq 4\sqrt{\log m})=O(m^{-8})$. So we only consider $0\leq \lambda\leq 4\sqrt{\log m}$. Let $N$ be a fixed positive integer and $\lambda_{i}=(i/N)\sqrt{\log m}$. The final screen level is selected by maximizing the number of rejections, i.e., $$\hat{\lambda}=(\hat{i}/N)\sqrt{\log m},\mbox{\quad where}\quad
\hat{i}=\argmax_{0\leq i\leq 4N}|\mathcal{R}_{\lambda_{i}}|.
$$
If there are several $i$ attain the maximum value, we
choose $\hat{i}$ to be the largest one among them.  Based on $\hat{\lambda}$, we can obtain $\mathcal{R}_{\hat{\lambda}}$ and the
final FDR control procedure is as follow.

\begin{center}
\begin{boxedminipage}{1.0\textwidth}
{\bf FDR control with US testing.}  For a target FDR $0<\alpha<1$, reject $H_{i0}$ if and only if $i\in\mathcal{R}_{\hat{\lambda}}$.
\end{boxedminipage}
\end{center}
\vspace{3mm}
The simulation
shows that the performance of the procedure is quite insensitive to the choice of $N$ when $N\geq 10$.

\section{Theoretical results}

\subsection{FDR and FDP control}

In this section, we state some theoretical results for the US testing procedure. Let $\mathcal{H}_{1}=\{1\leq i\leq m: \mu_{i,1}\neq\mu_{i,2}\}$. The following  conditions  are needed to establish the main results.\vspace{1mm}

\noindent{\bf (C1)}.
$|\mathcal{H}_{1}|=o(m)$ as $m\rightarrow\infty$.
\vspace{2mm}

\noindent{\bf (C2).} Assume that $\ep\exp(t_{0}|X_{i,j}-\mu_{i,j}|/\sigma_{i,j})\leq K_{1}$ for some $K_{1}>0$, $t_{0}>0$, all $1\leq i\leq m$ and $j=1,2$. Suppose that
$c_{1}\leq n_{1}/n_{2}\leq c_{2}$ and $c_{1}\leq \sigma^{2}_{i,1}/\sigma^{2}_{i,2}\leq c_{2}$ for some $c_{1},c_{2}>0$ and all $1\leq i\leq m$. The sample sizes satisfy
 $\min(n_{1},n_{2})\geq c(\log m)^{\zeta}$ for some $\zeta>5$ and $c>0$.\vspace{2mm}

In (C1),  we assume that the mean difference $\m_{1}-\m_{2}$ is sparse. The sparsity commonly arise from many applications such as the selection of differential expression genes. (C2) is a moment condition for populations which is regular in high-dimensional setting.
Let $\mathcal{M}_{i}$ be a subset of $\mathcal{H}_{0}$ such that $\{(X_{j,1},X_{j,2}), j\in \mathcal{M}_{i}\}$ is independent with $(X_{i,1},X_{i,2})$.

\begin{itemize}

\item[{\bf (C3).}] For every $i\in\mathcal{H}_{0}$, $|\mathcal{M}_{i}|\geq m_{0}-K$ for some $K>0$.

\end{itemize}

In (C3), for any $X_{i,1}$ and $X_{i,2}$, we allow $K$  variables which can be strongly correlated with them.
 Define
$$\mathcal{R}_{0,\lambda}=\Big{\{}i\in\mathcal{H}_{0}:~I\{|S_{i}|\geq \lambda,|T_{i}|\geq \hat{t}_{1,\lambda}\}=1\mbox{~or~}I\{|S_{i}|< \lambda,|T_{i}|\geq \hat{t}_{2,\lambda}\}=1\Big{\}}.$$
The FDP and FDR for the US  procedure are
\begin{eqnarray*}
FDP=\frac{|\mathcal{R}_{0,\hat{\lambda}}|}{\max(1,|\mathcal{R}_{\hat{\lambda}}|)}\quad\mbox{and}\quad FDR=\ep[FDP].
\end{eqnarray*}

\begin{theorem}\label{th1} Assume that (C2) and (C3) hold.
Suppose that
\begin{eqnarray}\label{c3}
|\mathcal{R}_{\hat{\lambda}}|\rightarrow\infty\quad\mbox{in probability}
\end{eqnarray}
as $m\rightarrow\infty$.
We have for any $\varepsilon>0$,
\begin{eqnarray}\label{r1}
\pr\Big{(}FDP\leq \alpha+\varepsilon\Big{)}\rightarrow 1
\end{eqnarray}
as $m\rightarrow\infty$. Consequently, limsup$_{m\rightarrow\infty}FDR\leq \alpha$.
\end{theorem}

Theorem \ref{th1} shows that the US procedure controls the FDR and FDP at level $\alpha$ asymptotically. We now discuss  condition
(\ref{c3}).
Actually, if the p-values $p_{j}$, $j\in\mathcal{H}_{i0}$, are i.i.d. $U(0,1)$ random variables, then Ferreira and Zwinderman (2006) prove that
 \begin{eqnarray}\label{c5}
\hat{R}_{BH}\rightarrow\infty\quad\mbox{in probability}
  \end{eqnarray}
  if and only if $FDP_{BH}\rightarrow \frac{m_{0}}{m}\alpha$ in probability, where   $FDP_{BH}$ is the false discovery proportion of the B-H method and $\hat{R}_{BH}$ is the number of rejections. So (\ref{c5}) is a sufficient and necessary condition for the FDP control of the B-H method.
 By the definition of the US procedure, $|\mathcal{R}_{0}|=\hat{R}_{BH}$, and hence (\ref{c5}) implies (\ref{c3}). Therefore, we conjecture that (\ref{c3}) is also a nearly necessary condition for the FDP control (\ref{r1}).
 A sufficient condition for (\ref{c3}) and (\ref{c5}) is
 \begin{eqnarray}
 Card\Big{\{}1\leq i\leq m:~ \frac{|\mu_{i,1}-\mu_{i,2}|}{\sqrt{\sigma^{2}_{i,1}/n_{1}+\sigma^{2}_{i,2}/n_{2}}}\geq\theta\sqrt{\log m}\Big{\}}\rightarrow\infty
 \mbox{\quad for some $\theta>\sqrt{2}$,}
 \end{eqnarray}
  which is quite mild.\vspace{2mm}

\subsection{Power comparison}

In this section, we compare the US procedure to the B-H method.
Define the power of the B-H method by
\begin{eqnarray*}
\text{power}_{BH}=\frac{\sum_{i\in\mathcal{H}_{1}}\I\{p_{i}\leq p_{(\hat{k})}\}}{m_{1}},
\end{eqnarray*}
where $p_{i}=2-2\Psi(|T_{i}|)$ and $m_{1}=|\mathcal{H}_{1}|$.
The power of the US procedure is defined by
\begin{eqnarray}\label{po}
\text{power}_{US}=\frac{|\mathcal{R}_{\hat{\lambda}}|-|\mathcal{R}_{0,\hat{\lambda}}|}{m_{1}}.
\end{eqnarray}
We first show that the US procedure can be at least as powerful as the B-H method asymptotically without requiring any sparsity on $\m_{1}$ and $\m_{2}$.

\begin{theorem}\label{th5} Assume $m_{1}\rightarrow\infty$ and (C1)-(C3) hold. Then we have
\begin{eqnarray*}
\text{power}_{US}\geq \text{power}_{BH}+o_{\pr}(1)
\end{eqnarray*}
for some $o_{\pr}(1)$ as $m\rightarrow\infty$.
\end{theorem}

 The condition $m_{1}\rightarrow\infty$ is a necessary condition for the FDP control of the B-H method; see Proposition 2.1 in Liu and Shao (2014).
When $m_{1}$ is fixed as $m\rightarrow\infty$, the  true FDPs of the B-H method and the US procedure will suffer from drastic fluctuations, and hence in this case we do not consider the power comparison  under the FDP control. On the other hand,
theoretical derivations for the power comparison under FDR control
are typically more complicated when $m_{1}$ is fixed. We leave this as a future work.

We next investigate the power of the B-H method.
Assume that
 \begin{eqnarray}\label{c4}
 \frac{|\mu_{i,1}-\mu_{i,2}|}{\sqrt{\sigma^{2}_{i,1}/n_{1}+\sigma^{2}_{i,2}/n_{2}}}=\theta\sqrt{\log m},~i\in\mathcal{H}_{1}
 \end{eqnarray}
 for some $\theta>0$.
The number of signals is assumed to be
\begin{eqnarray}\label{c6}
|\mathcal{H}_{1}|=p^{\beta} \mbox{\quad for some $0<\beta<1$.}
 \end{eqnarray}
 We have the following theorem for power$_{BH}$.

\begin{theorem}\label{th3}  Suppose that (C2) and (C3) hold. If $0<\theta<\sqrt{2(1-\beta)}$, then we have $\text{power}_{BH}\rightarrow 0$ in probability as $m\rightarrow\infty$.
If $\theta>\sqrt{2(1-\beta)}$, then $\text{power}_{BH}\rightarrow 1$ in probability as $m\rightarrow\infty$.
\end{theorem}

Theorem  \ref{th3} reveals an interesting critical phenomenon for the B-H method. It indicates that when the size of signals satisfies $0<\theta<\sqrt{2(1-\beta)}$, then the B-H method is unable to detect most of signals.
On the other hand, if $\theta>\sqrt{2(1-\beta)}$, then the power of the B-H method converges to one. In this case, by Theorem \ref{th5}, power$_{US}$ will also converges to one in probability.

We shall show that, when $0<\theta<\sqrt{2(1-\beta)}$, power$_{US}$ can converge to one for a wide class of $\m_{1}$ and $\m_{2}$.
To this end, assume that  $\m_{1}$ and $\m_{2}$ satisfy
\begin{eqnarray}\label{c7}
Card\Big{\{}i\in\mathcal{H}_{0}:~\sqrt{\frac{n_{1}}{\sigma^{2}_{i,1}(1+\frac{n_{2}\sigma^{2}_{i,1}}{n_{1}\sigma^{2}_{i,2}})}}
\Big{|}\mu_{i,1}+\frac{n_{2}\sigma^{2}_{i,1}}{n_{1}\sigma^{2}_{i,2}}\mu_{i,2}\Big{|}\geq h\sqrt{\log m}\Big{\}}=O(m^{\gamma})
\end{eqnarray}
and
\begin{eqnarray}\label{c8}
Card\Big{\{}i\in\mathcal{H}_{1}:~\sqrt{\frac{n_{1}}{\sigma^{2}_{i,1}(1+\frac{n_{2}\sigma^{2}_{i,1}}{n_{1}\sigma^{2}_{i,2}})}}
\Big{|}\mu_{i,1}+\frac{n_{2}\sigma^{2}_{i,1}}{n_{1}\sigma^{2}_{i,2}}\mu_{i,2}\Big{|}\geq \kappa\sqrt{\log m}\Big{\}}\geq \rho|\mathcal{H}_{1}|
\end{eqnarray}
for some $0\leq\gamma\leq 1$, $0<h\leq 2$, $0<\rho\leq 1$ and $\kappa>h+\sqrt{2(1-\beta)}$. (\ref{c7}) is an asymptotic sparsity condition on $\m_{1}$ and $\m_{2}$. It is quite mild as $m^{\gamma}$ elements can be arbitrarily large and the other elements  can be of the order of $\sqrt{(\log m)/n}$.
Condition (\ref{c8}) is needed to ensure that signals in $\mathcal{H}_{1}$ can be screened into the first family of hypotheses by $S_{i}$.

\begin{theorem}\label{th4} Suppose that  (C2), (C3), (\ref{c4}) and (\ref{c6}) hold.

(i). If $\theta>\sqrt{2(1-\beta)}$, then $\text{power}_{US}\rightarrow 1$ in probability as $m\rightarrow\infty$.

(ii). Assume that (\ref{c7}) and (\ref{c8}) hold. Let  $\theta>\sqrt{\max(0,2\gamma-2\beta)}$ and  $N\geq 10/\min(1-\beta,\theta^{2}/4)$. We have
$\pr(\text{power}_{US}\geq \rho-\varepsilon)\rightarrow 1$ for any $\varepsilon>0$ as $m\rightarrow\infty$.

(iii). We have $\pr(FDP\leq \alpha+\varepsilon)\rightarrow 1$ for any $\varepsilon>0$ as $m\rightarrow\infty$.
\end{theorem}

Theorem \ref{th4} indicates that, if $\m_{1}$ and $\m_{2}$ satisfy (\ref{c7}) and (\ref{c8}), then power$_{US}$ can be much larger than power$_{BH}$.
In particular, the power of US procedure converges to one when $\rho=1$ and $\theta>\sqrt{\max(0,2\gamma-2\beta)}$.
In contrast, if $\sqrt{\max(0,2\gamma-2\beta)}<\theta<\sqrt{2(1-\beta)}$, power$_{BH}$ converges to zero.\vspace{2mm}

\noindent{\bf Remark.} Condition (\ref{c7}) is quite mild. For example, in ultra-high dimensional setting $\log m\geq c\max(n_{1},n_{2})$ for some $c>0$,
 all of components of $\m_{1}$ and $\m_{2}$ can be bounded away from zero.  In this case, (\ref{c7}) essentially is not an asymptotic sparsity condition.  In (C2),
 we require    $\min(n_{1},n_{2})\geq c(\log m)^{\zeta}$ for some $\zeta>5$. However, this condition is only used to ensure that the sample variances
and  null distribution of $T_{i}$ are close to the population variances and $\Psi(t)$, respectively. In the ideal case
 that $\X_{1}$ and $\X_{2}$ are multivariate normal random vectors with known variances,
 we can use $T_{0i}=(\bar{X}_{i,1}-\bar{X}_{i,2})/\sqrt{\sigma^{2}_{i,1}/n_{1}+\sigma^{2}_{i,2}/n_{2}}$ as a test statistic with $N(0,1)$ null distribution and $S_{0i}$ as a screening statistic. Then all theorems hold without (C2). In this case, (\ref{c7})  allows non-sparse $\m_{1}$ and $\m_{2}$   in ultra-high dimensional setting. Although $\X_{1}$ and $\X_{2}$ may be non-Gaussian, we will show by numerical studies in Section 4 that the US procedure indeed outperforms the B-H method
 for non-sparse mean vectors when $m$ is large.

\section{Numerical results}

In this section, we conduct numerical simulations and examine the performance of the US procedure.
Let
\begin{eqnarray*}
\X_{1}=\m_{1}+\va_{1}-\ep\va_{1}\mbox{\quad and\quad}\X_{2}=\m_{1}+\va_{2}-\ep\va_{2},
\end{eqnarray*}
where $\va_{1}=(\varepsilon_{1,1},\ldots,\varepsilon_{m,1})$ and $\va_{2}=(\varepsilon_{1,2},\ldots,\varepsilon_{m,2})$ are independent  random vectors.\vspace{2mm}

{\bf Model 1.}  Let
$
\mu_{i,1}=3\sqrt{\frac{\log m}{n_{1}}}
$ and $\mu_{i,2}=2\sqrt{\frac{\log m}{n_{2}}}$
for $1\leq i\leq m_{1}$;  $\mu_{i,1}=\mu_{i,2}=0$ for $m_{1}+1\leq i\leq m$.\vspace{2mm}

{\bf Model 2.}  Let
$
\mu_{i,1}=2\sqrt{\frac{\log m}{n_{1}}}
$
for $1\leq i\leq m_{1}$; $\mu_{i,2}=\sqrt{\frac{\log m}{n_{2}}}$ for $1\leq i\leq
[m_{1}/2]$; $\mu_{i,2}=-0.5\sqrt{\frac{\log m}{n_{2}}}$ for $[m_{1}/2]+1\leq i\leq
m_{1}$; $\mu_{i,1}=\mu_{i,2}=0$ for  $m_{1}+1\leq i\leq m$.\vspace{2mm}

{\bf Model 3.}  Let
$
\mu_{i,1}=3\sqrt{\frac{\log m}{n_{1}}}
$ and $\mu_{i,2}=2\sqrt{\frac{\log m}{n_{2}}}$
for $1\leq i\leq m_{1}$;  $\mu_{i,1}=\mu_{i,2}=(i/m)\sqrt{(\log m)/n_{1}}$ for $m_{1}+1\leq i\leq m$. \vspace{2mm}

{\bf Model 4.}  Let
$
\mu_{i,1}=3\sqrt{\frac{\log m}{n_{1}}}
$ and $\mu_{i,2}=2\sqrt{\frac{\log m}{n_{2}}}$
for $1\leq i\leq m_{1}$; $\mu_{i,1}=\mu_{i,2}=1$ for $m_{1}+1\leq i\leq m_{1}+[\sqrt{m}]$; $\mu_{i,1}=\mu_{i,2}=0.2$ for $m_{1}+[\sqrt{m}]+1\leq i\leq m$. \vspace{2mm}

 In Models 1 and 2, $\m_{1}$ and $\m_{2}$ are exactly sparse, and they are asymptotically sparse in Model 3. In Model 4, $\m_{1}$ and $\m_{2}$
  are non-sparse vectors.
 We take $n_{1}=n_{2}=100$, $p=2000$ and $m_{1}=[\sqrt{m}]$. We consider the US procedure for {\em equal variances case} and {\em unequal variances case}.
 In the first case, we let $\varepsilon_{i,1}$ and $\varepsilon_{i,2}$, $1\leq i\leq m$, be i.i.d. $N(0,1)$ variables. In the second case,
 $\varepsilon_{i,1}\sim N(0,0.5)$ and $\varepsilon_{i,2}\sim N(0,1)$. We also carry out simulation studies for $t$-distributed errors and simulation results are stated in the supplementary material Liu (2014).

  The simulation is replicated 500 times and $N$ in $\hat{\lambda}$ is taken to be 10. Extensive simulations indicate that the performance of the US procedure is quite insensitive to the choice of $N$ when $N\geq 10$. We calculate the empirical powers of the US procedure and the B-H method by
\begin{eqnarray*}
\text{power}_{US}=\frac{1}{500}\sum_{i=1}^{500}\text{power}_{US,i}\mbox{~~and~~}\text{power}_{BH}=\frac{1}{500}\sum_{i=1}^{500}\text{power}_{BH,i},
\end{eqnarray*}
where $\text{power}_{US,i}$ and $\text{power}_{BH,i}$ are the powers of the US procedure and the B-H method in the $i$-th replication, respectively. Also,
 the empirical FDRs are obtained by the average of all FDPs in 500 replications:
  \begin{eqnarray*}
\text{eFDR}_{US}=\frac{1}{500}\sum_{i=1}^{500}\text{FDP}_{US,i}\mbox{~~and~~}\text{eFDR}_{BH}=\frac{1}{500}\sum_{i=1}^{500}\text{FDP}_{BH,i},
\end{eqnarray*}
  The target FDR is taken to be $\alpha=i/20$, $1\leq i\leq 20$ so that we can compare the US procedure and the B-H method along a series of $\alpha$. The empirical powers power$_{US}$ and power$_{BH}$ are plotted in Figures 1 and 2 for
 all $\alpha=i/20$, $1\leq i\leq 20$.
From Figures 1 and 2, we can see that the US procedure  has much more statistical power than the B-H method on all four models.
For example, in Model 1, power$_{BH}$ is below $0.1$ for $\alpha\leq 0.3$,
while power$_{US}$ grows from 0.3 to 0.7 as $\alpha$ grows from $0.05$ to $0.3$. Similar phenomenon can be observed for other models. In particular, for the non-sparse Model 4, the US procedure is still significantly more powerful than the B-H method.

To examine the performance of FDR control,  we consider the ratio between the empirical FDR and the target FDR $\alpha$.
 The values $\text{eFDR}/\alpha$ are plotted in Figures 3 and 4. We can see that the ratios for Models 1-4 are always close to or smaller than  1.
 Hence, the US procedure can control the FDR effectively while having more power than the B-H method. Note that in many cases, the FDRs of the US procedure are  smaller than
 $\alpha$. The possible reason is that $\hat{m}_{1,\lambda}$ overestimates $\hat{m}^{o}_{1,\lambda}$ as $\mathcal{B}_{1}$ usually contains more true alternatives than true nulls. So the FDR in the first family of hypotheses will be smaller than $\alpha$. Overall,  the US procedure is much more powerful than the B-H method, and interestingly, it has smaller FDRs when $\alpha\leq 0.5$.

We next examine the performance of other seemingly natural screening methods including the square type screening statistics and  maximum type screening statistics. Let $T_{i,1}$ and $T_{i,2}$ be one-sample {\em Student}'s statistics $T_{i,1}=\sqrt{n_{1}}\bar{X}_{i,1}/\hat{\sigma}_{i,1}$ and
$T_{i,2}=\sqrt{n_{2}}\bar{X}_{i,2}/\hat{\sigma}_{i,2}$. The square type screening and maximum type screening use
$SS_{i}=\sqrt{T^{2}_{i,1}+T^{2}_{i,2}}$  and  $MS_{i}=\max(|T_{i,1}|,|T_{i,2}|)$ as screening statistics, respectively. Now we replace $S_{i}$ in
the US procedure by $SS_{i}$ and $MS_{i}$ and replicate the above numerical studies for Model 4. The screen level $\lambda$ is chosen to be $\hat{\lambda}$ or $\sqrt{2\log m}$. The ratios eFDR$/\alpha$ are plotted
in Figure 5. We can see that neither the square type screening nor the maximum type screening controls the FDR. The reason is that $SS_{i}$ and $MS_{i}$
are correlated with $T_{i}$ so that p-values are no longer $U(0,1)$ after screening.

Finally, we show that {\em testing after screening} with sample splitting may loss much statistical power. To see this, we consider the following model.

{\bf Model 5.}  Let
$
\mu_{i,1}=1.5\sqrt{\frac{\log m}{n_{1}}}
$ and $\mu_{i,2}=-0.5\sqrt{\frac{\log m}{n_{1}}}$
for $1\leq i\leq [\sqrt{m}]$;  $\mu_{i,1}=\mu_{i,2}=0$ for $[\sqrt{m}]+1\leq i\leq m$.\vspace{2mm}

In the screening stage, we use 50 samples to construct screening statistics $SS_{i}$ and $MS_{i}$. The two-sample {\em Student}'s statistics $T_{i}$ are constructed from  the remaining 50 samples. The thresholding level in  screening stage is chosen by the same way as $\hat{\lambda}$.
We plot power curves in Figure 6 for $SS_{i}$ screening and $MS_{i}$ screening. It can be observed that the sample splitting method results in a significant power loss, comparing to the B-H method and the US procedure.

\section{Discussion}

In this article, we consider the FDR/FDP control for two-sample multiple $t$ tests. The proposed US procedure is shown to be more powerful than the classical B-H method. There are several possible extensions.

In the  setting of dense signals, it is well known that an accurate estimator for the number of true null hypotheses can help improve the power of the B-H method; see Storey, et al. (2004).  The latter paper develops an estimator $\hat{m}_{0}$ for $m_{0}$ and then incorporates it into the B-H method. Similarly,  we can develop some accurate estimates for $\hat{m}^{o}_{1,\lambda}$ and $\hat{m}^{o}_{2,\lambda}$ to replace $\hat{m}_{1,\lambda}$ and $\hat{m}_{2,\lambda}$. The power of the US procedure is expected to be improved
in this way and theoretical study is left for future work.

Controlling the FDR under dependence is an important and challenging topic. Many procedures for FDR control under various dependence frameworks have been developed. Leek and Storey (2008) consider a general framework for multiple tests in the presence of arbitrarily strong dependence. Friguet, et al. (2009)
consider the FDR control under the factor model assumption. Fan, et al. (2012) estimate the false discovery proportion under arbitrary covariance dependence.
It would be interesting to study the US procedure under these dependence settings.

The uncorrelated screening technique can be extended to other related two-sample testing problems. For example, consider the two sample correlation testing problem
$H_{0ij}:$ $\rho_{ij1}=\rho_{ij2}$, $1\leq i<j\leq m$, where $\R_{1}=(\rho_{ij1})_{1\leq i,j\leq m}$ and $\R_{2}=(\rho_{ij2})_{1\leq i,j\leq m}$ are two correlation matrices. The correlation matrix is often assumed to be (asymptotically) sparse; see Bickel and Levina (2008). The uncorrelated screening technique
can be applied in this problem. Similarly, it can be applied in  two sample partial correlation testing problem $H_{0ij}:$ $\rho^{'}_{ij1}=\rho^{'}_{ij2}$, $1\leq i<j\leq m$, where $\rho^{'}_{ij1}$ and $\rho^{'}_{ij2}$ denote the partial correlation coefficients which are closely related to Gaussian graphical models (GGM). In GGM estimation, it is common to assume the sparsity on the partial correlation coefficients; see Liu (2013).

\begin{sidewaysfigure}
\begin{center}
\subfloat[][Model 1]
{\includegraphics[width=0.25\textwidth]{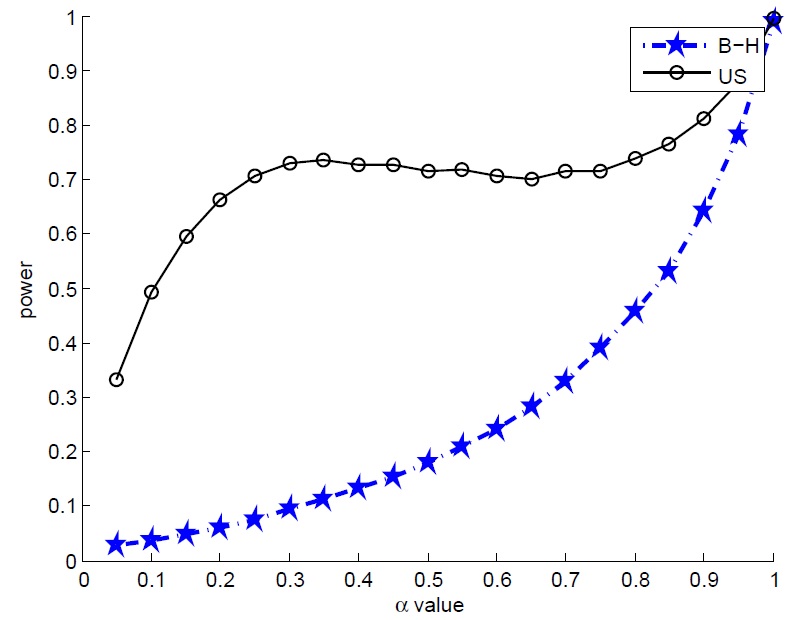} }
\subfloat[][Model 2]
{\includegraphics[width=0.25\textwidth]{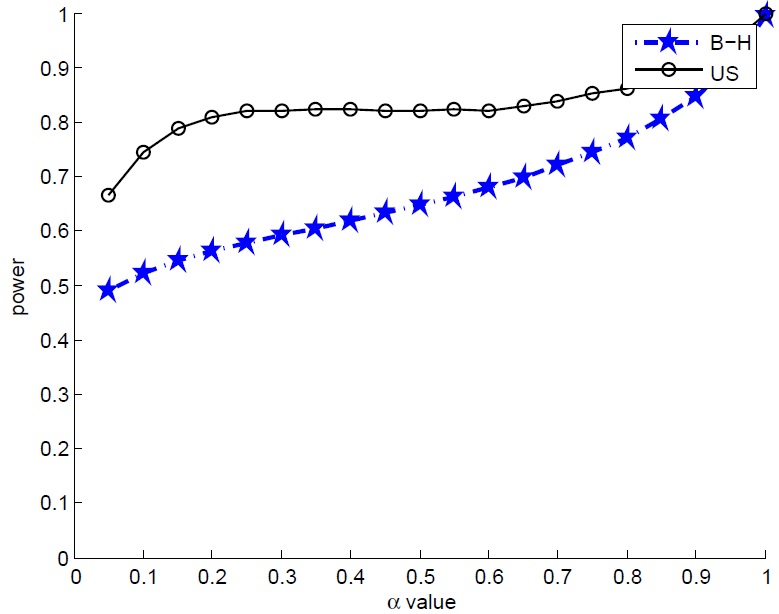} }
\subfloat[][Model 3]
{\includegraphics[width=0.25\textwidth]{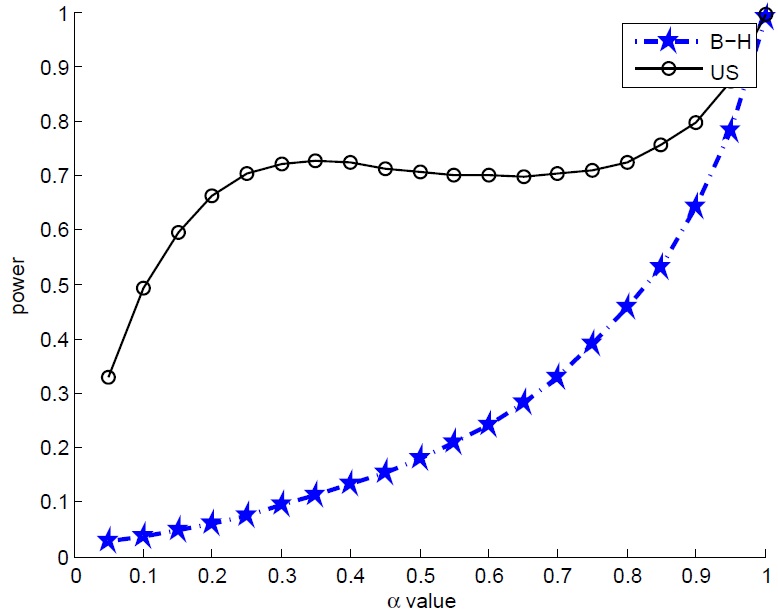} }
\subfloat[][Model 4]
{\includegraphics[width=0.25\textwidth]{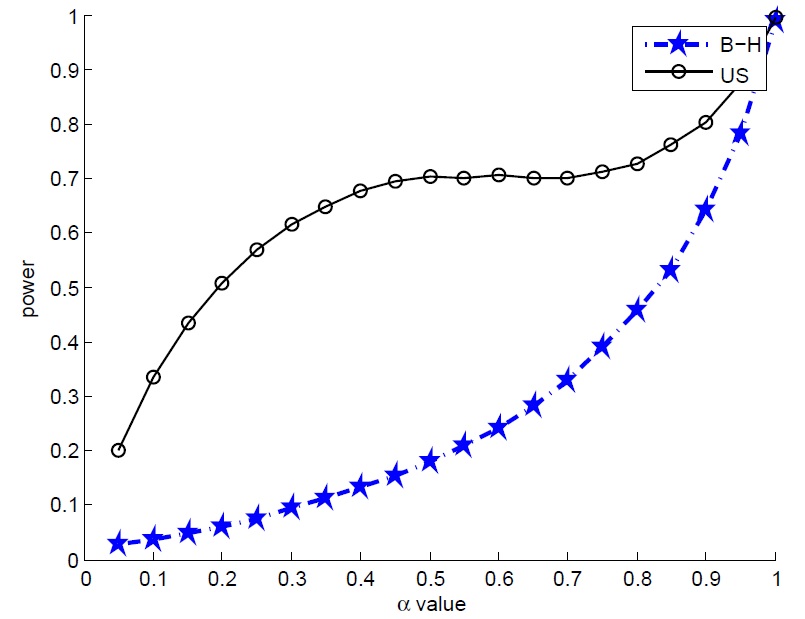} }
\end{center}
 \caption{Equal variances. The x-axis denotes the $\alpha$ value and the y-axis denotes the power.}
 \subfloat[][Model 1]
{\includegraphics[width=0.25\textwidth]{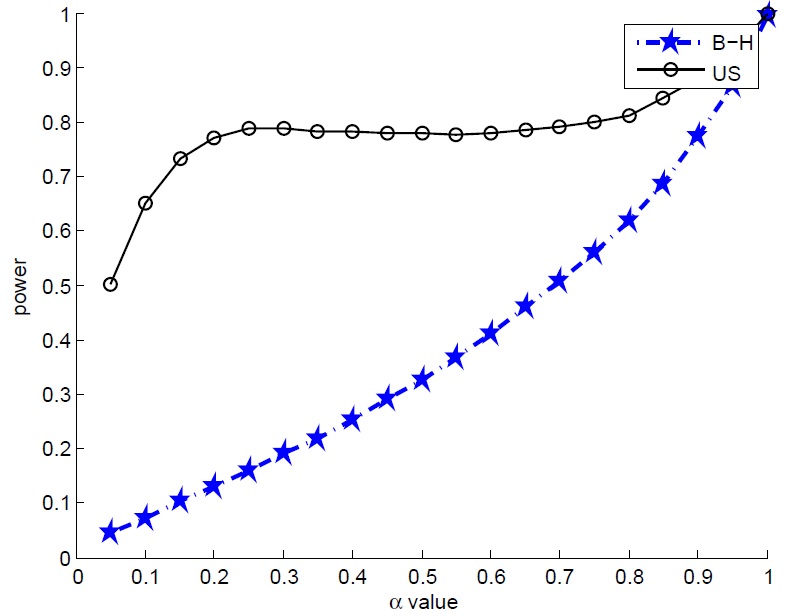} }
\subfloat[][Model 2]
{\includegraphics[width=0.25\textwidth]{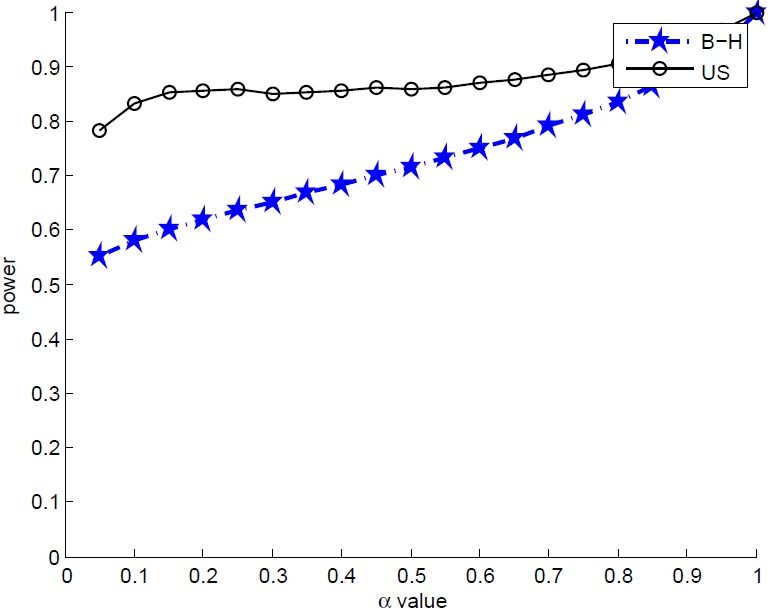} }
\subfloat[][Model 3]
{\includegraphics[width=0.25\textwidth]{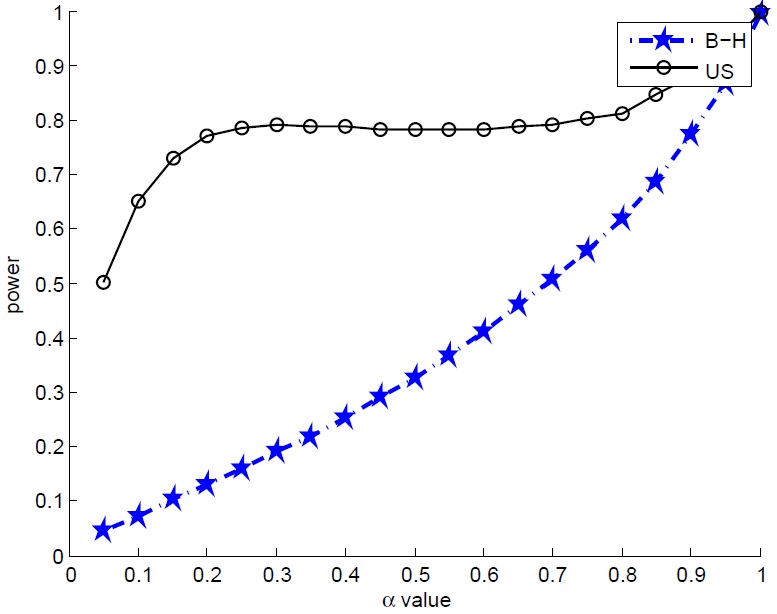} }
\subfloat[][Model 4]
{\includegraphics[width=0.25\textwidth]{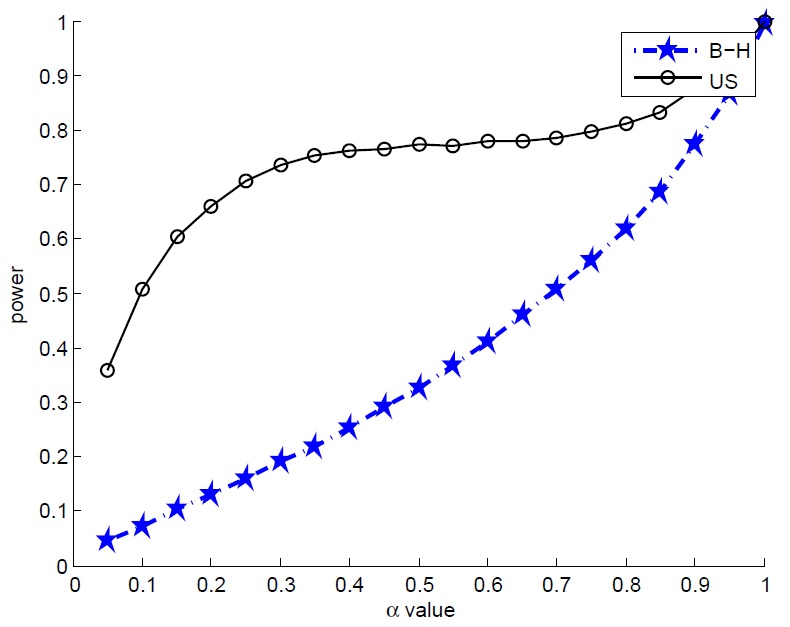} }
\caption{Unequal variances. The x-axis denotes the $\alpha$ value and the y-axis denotes the power.}
 \end{sidewaysfigure}

\begin{sidewaysfigure}
\begin{center}
\subfloat[][Model 1]
{\includegraphics[width=0.25\textwidth]{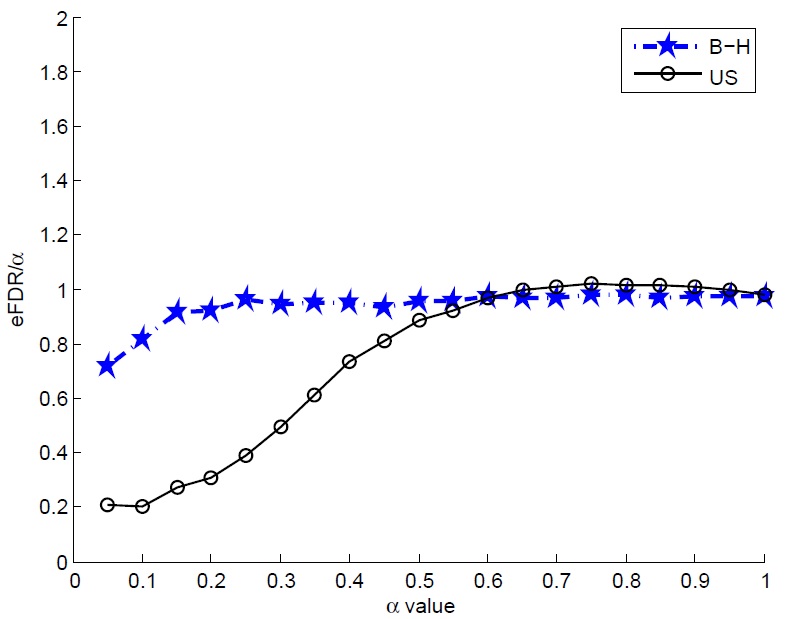} }
\subfloat[][Model 2]
{\includegraphics[width=0.25\textwidth]{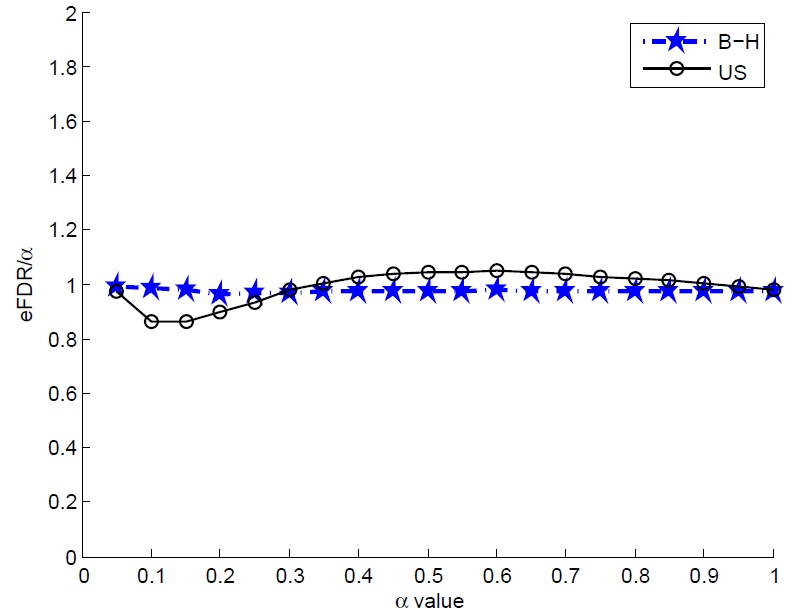} }
\subfloat[][Model 3]
{\includegraphics[width=0.25\textwidth]{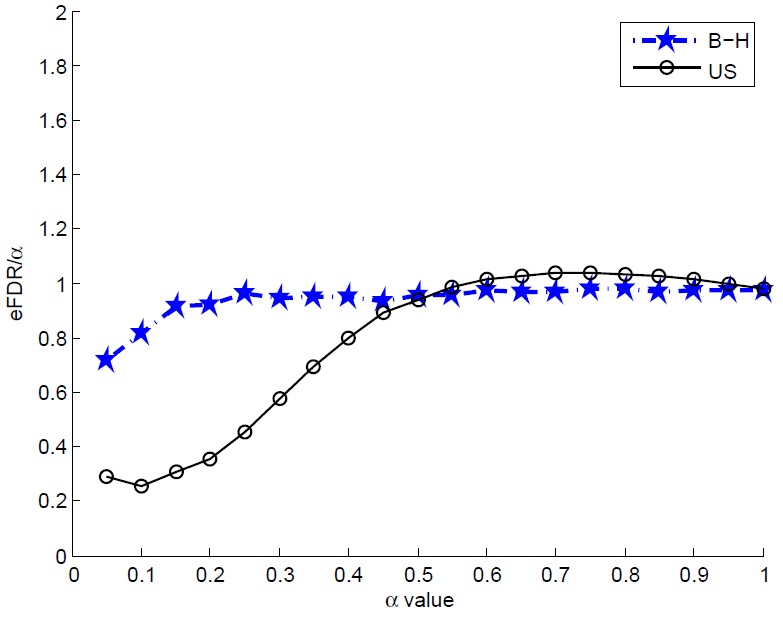} }
\subfloat[][Model 4]
{\includegraphics[width=0.25\textwidth]{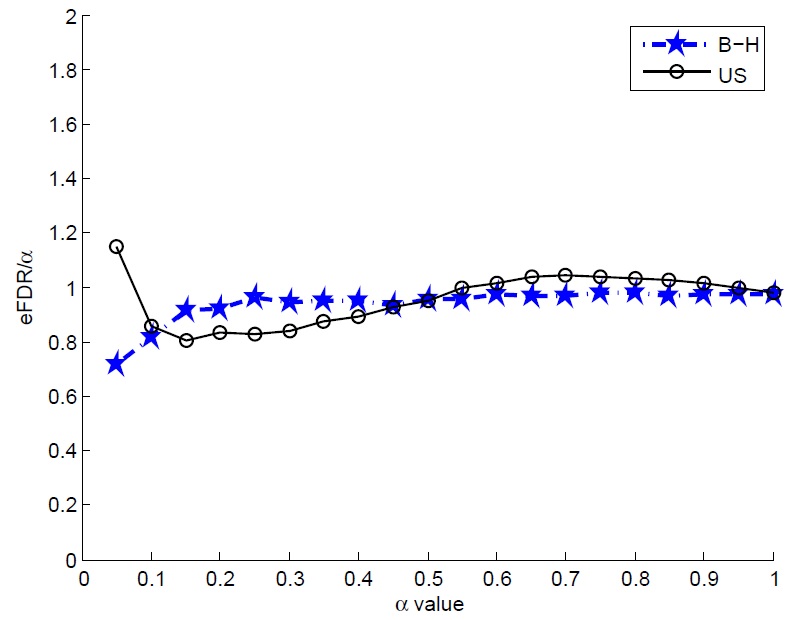} }
\end{center}
 \caption{Equal variances. The x-axis denotes the $\alpha$ value and the y-axis denotes eFDR$/\alpha$.}
 \subfloat[][Model 1]
{\includegraphics[width=0.25\textwidth]{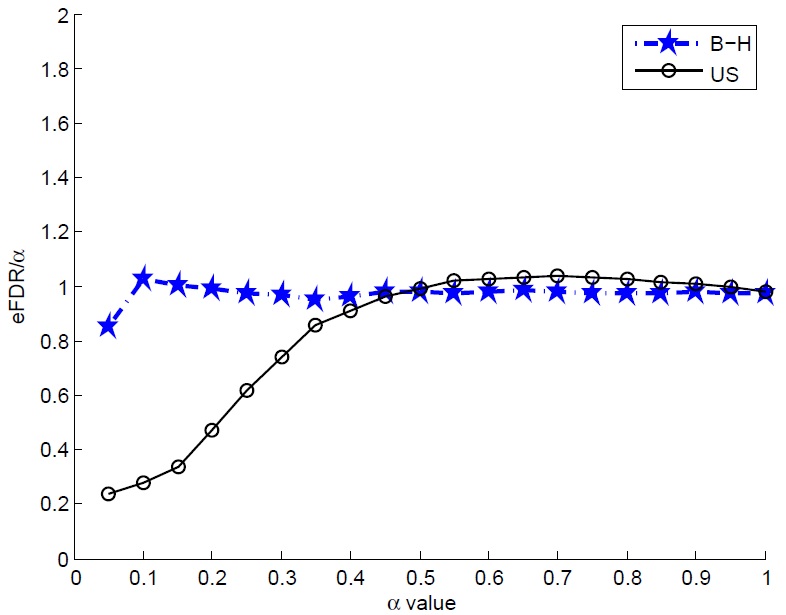} }
\subfloat[][Model 2]
{\includegraphics[width=0.25\textwidth]{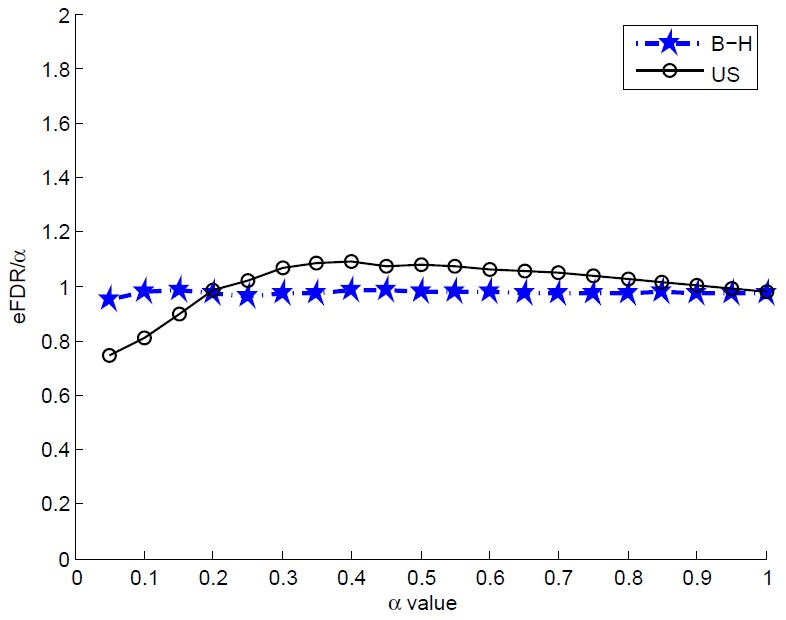} }
\subfloat[][Model 3]
{\includegraphics[width=0.25\textwidth]{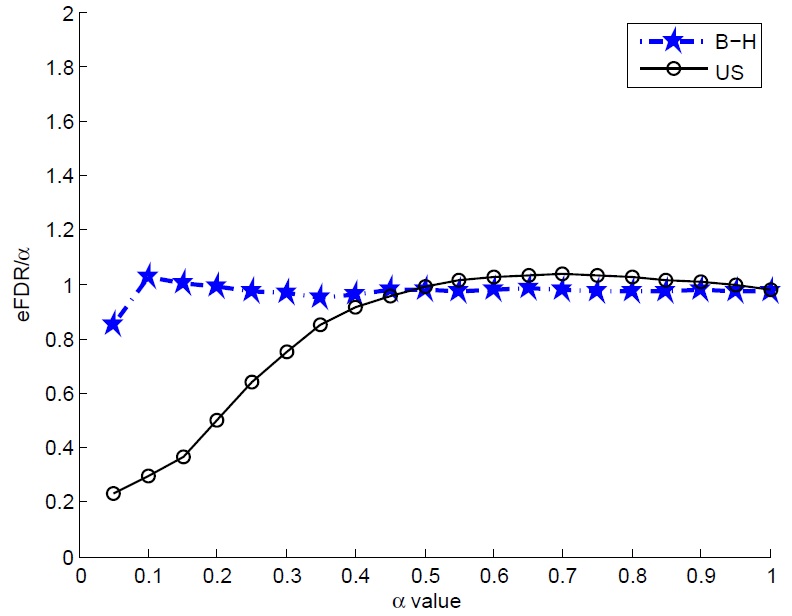} }
\subfloat[][Model 4]
{\includegraphics[width=0.25\textwidth]{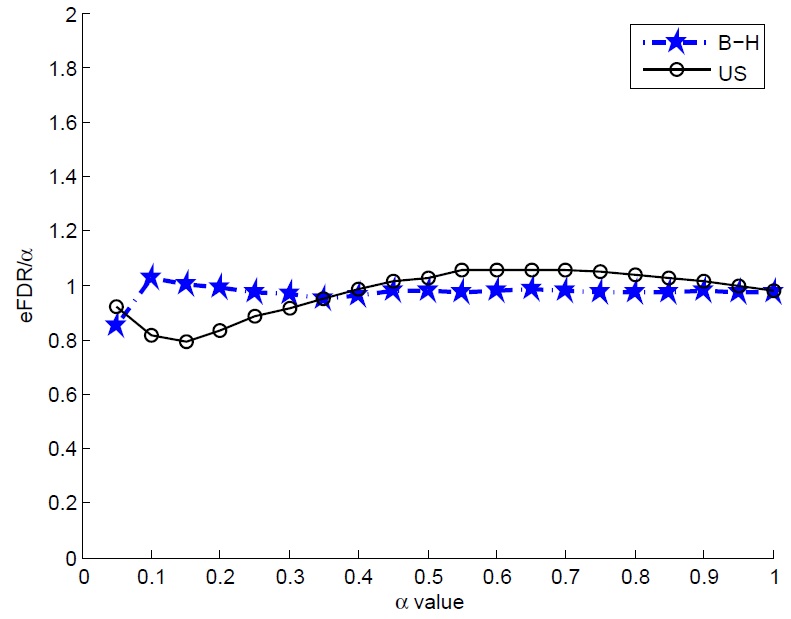} }
\caption{Unequal variances. The x-axis denotes the $\alpha$ value and the y-axis denotes eFDR$/\alpha$.}
 \end{sidewaysfigure}


\begin{figure}[htbp]
\begin{center}
\subfloat[][ $\lambda=\hat{\lambda}$.]
{\includegraphics[width=0.4\textwidth]{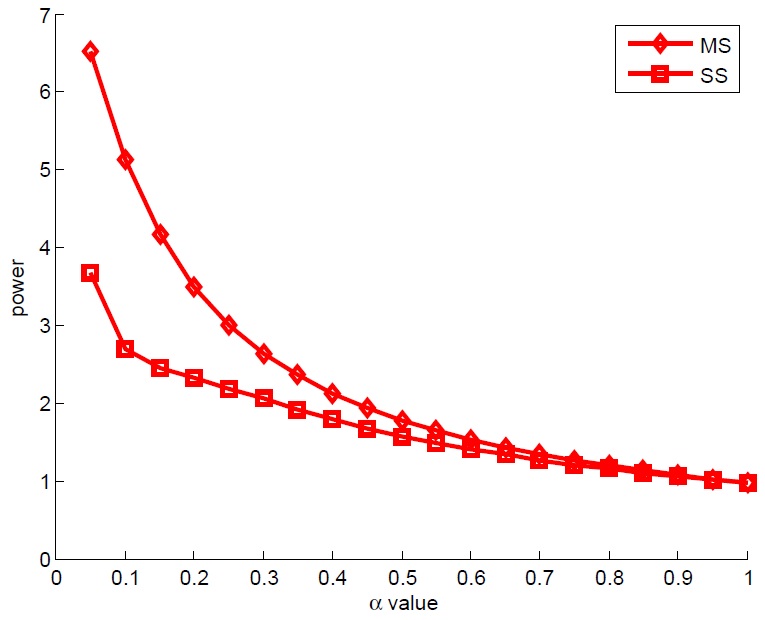} }
\subfloat[][ $\lambda=\sqrt{2\log m}$.]
{\includegraphics[width=0.4\textwidth]{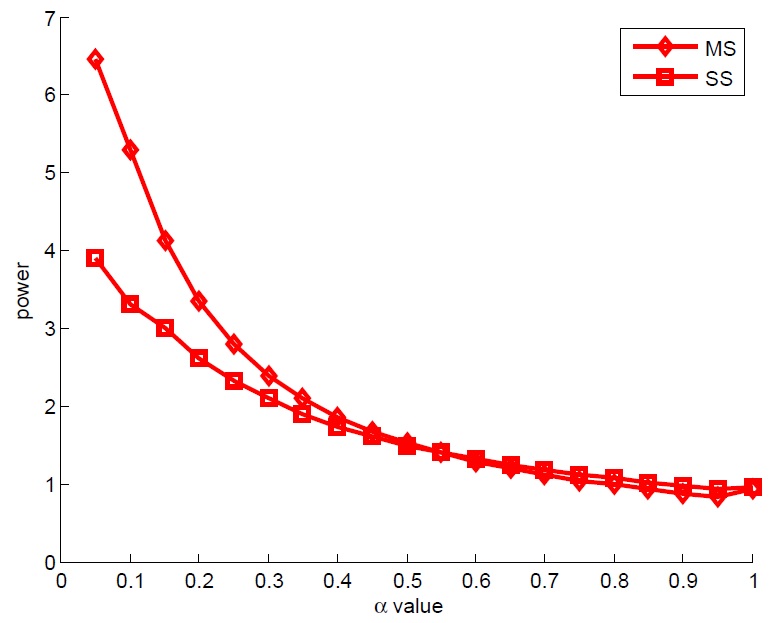} }
\end{center}
 \caption{Model 4 and equal variances. The x-axis denotes the $\alpha$ value and the y-axis denotes eFDR$/\alpha$ by screening with $SS_{i}$ and $MS_{i}$. }
\end{figure}

\begin{figure}[htbp]
\begin{center}
{\includegraphics[width=0.4\textwidth]{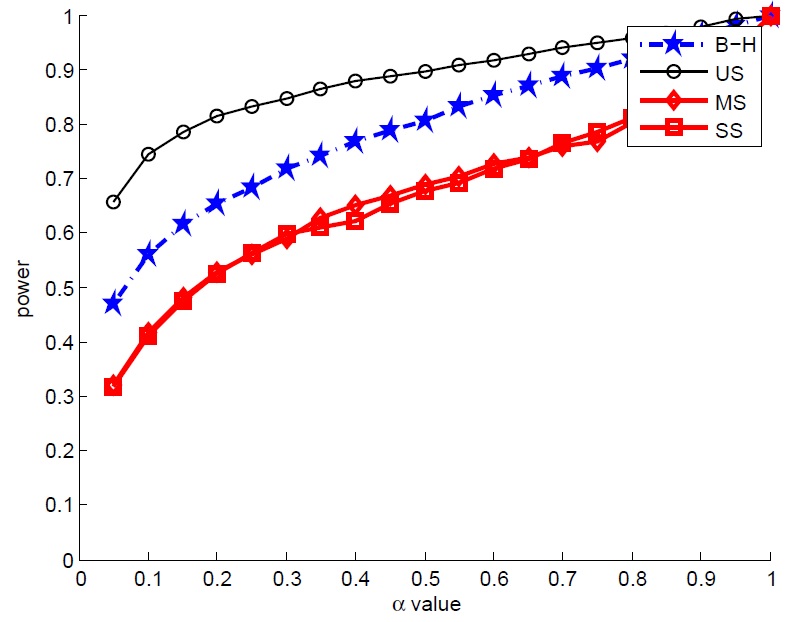} }
\end{center}
 \caption{Model 5 and equal variances. The x-axis denotes the $\alpha$ value and the y-axis denotes the powers of the US procedure, the B-H procedure and
the sample splitting method with MS$_{i}$ and SS$_{i}$ screening.}
\end{figure}

\section{Proof of main results}

We only prove the main results for Case II, variances $\sigma^{2}_{i,1}$ and $\sigma^{2}_{i,2}$ are not necessary equal because the proof for Case I is quite similar.

\subsection{Proof of Theorem \ref{th1}}

Let $c_{m}\rightarrow\infty$ such that $\pr(|\mathcal{R}_{\hat{\lambda}}|\geq c_{m})\rightarrow 1$ as $m\rightarrow\infty$. Define
\begin{eqnarray*}
m^{o}_{1,\lambda}=\sum_{i\in\mathcal{H}_{0}}\pr(|N(0,1)+h_{i}|\geq \lambda)\mbox{\quad and\quad} m^{o}_{2,\lambda}=\sum_{i\in\mathcal{H}_{0}}\pr(|N(0,1)+h_{i}|< \lambda),
\end{eqnarray*}
where
 \begin{eqnarray*}
 h_{i}=\sqrt{\frac{n_{1}}{\sigma^{2}_{i,1}(1+\frac{n_{2}\sigma^{2}_{i,1}}{n_{1}\sigma^{2}_{i,2}})}}\Big{(}\mu_{i,1}+\frac{n_{2}\sigma^{2}_{i,1}}{n_{1}\sigma^{2}_{i,2}}\mu_{i,2}\Big{)}.
 \end{eqnarray*}
 We first prove that for any  $b_{m}\rightarrow\infty$, $\varepsilon>0$ and $0\leq \lambda\leq 4\sqrt{\log m}$,
\begin{eqnarray}\label{a1}
\pr\Big{(}\sup_{0\leq t\leq G^{-1}(b_{m}/m^{o}_{1,\lambda})}\Big{|}\frac{\sum_{i\in\mathcal{H}_{0}}I\{|S_{i}|\geq \lambda,|T_{i}|\geq t\}}{m^{o}_{1,\lambda}G(t)}-1\Big{|}\geq \varepsilon\Big{)}\rightarrow 0
\end{eqnarray}
as $m\rightarrow\infty$,  where $G(t)=2-2\Psi(t)$ and $\sup_{0\leq t\leq G^{-1}(b_{m}/m^{o}_{1,\lambda})}(\cdot)=0$ if $b_{m}>m^{o}_{1,\lambda}$.
Note that we only need to consider the case $m^{o}_{1,\lambda}\geq b_{m}$. By (\ref{c9}), (\ref{lee4}), (\ref{le5-5}) and the proof of Lemma 6.3 in Liu (2013), it suffices to prove for any $\varepsilon>0$,
\begin{eqnarray}\label{p1}
\int_{0}^{G^{-1}(b_{m}/m^{o}_{1,\lambda})}\pr\Big{(}\Big{|}\frac{\sum_{i\in\mathcal{H}_{0}}f_{i}(\lambda,t)}{m^{o}_{1,\lambda}G(t)}\Big{|}\geq \varepsilon\Big{)}I\{m^{o}_{1,\lambda}\geq b_{m}\}dt=o(v_{m})
\end{eqnarray}
for some $v_{m}=o(1/\sqrt{\log m^{o}_{1,\lambda}})$ and
\begin{eqnarray}\label{p2}
\sup_{0\leq t\leq G^{-1}(b_{m}/m^{o}_{1,\lambda})}\pr\Big{(}\Big{|}\frac{\sum_{i\in\mathcal{H}_{0}}f_{i}(\lambda,t)}{m^{o}_{1,\lambda}G(t)}-1\Big{|}\geq \varepsilon\Big{)}I\{m^{o}_{1,\lambda}\geq b_{m}\}=o(1)
\end{eqnarray}
as $m\rightarrow\infty$, where $\log(x)=\ln(\max(x,e))$ and
$$f_{i}(\lambda,t)=I\{|S_{i}|\geq \lambda,|T_{i}|\geq t\}-\pr(|S_{i}|\geq \lambda,|T_{i}|\geq t).$$
By Lemma \ref{le4} and (C3), we have for any $M>0$,
\begin{eqnarray*}
\ep (f_{i}(\lambda,t))^{2}\leq C\{m^{o}_{1,\lambda}G(t)+m^{-M}\}
\end{eqnarray*}
uniformly in $0\leq t\leq 4\sqrt{\log m}$, $0\leq \lambda\leq 4\sqrt{\log m}$ and $i\in\mathcal{H}_{0}$. This proves (\ref{p2}). Note that
\begin{eqnarray*}
\int_{0}^{G^{-1}(b_{m}/m^{o}_{1,\lambda})}\frac{1}{m^{o}_{1,\lambda}G(t)}dt\leq \frac{C}{b_{m}\sqrt{\log(m^{o}_{1,\lambda}/b_{m})}}.
\end{eqnarray*}
Thus, we have (\ref{p1}).
By Lemma \ref{le5}, we have
\begin{eqnarray}\label{a1-1-1}
\pr\Big{(}\sup_{0\leq t\leq G^{-1}(b_{m}/m^{o}_{1,\lambda})}\Big{|}\frac{\sum_{i\in\mathcal{H}_{0}}I\{|S_{i}|\geq \lambda,|T_{i}|\geq t\}}{\hat{m}^{o}_{1,\lambda}G(t)}-1\Big{|}\geq \varepsilon\Big{)}\rightarrow 0
\end{eqnarray}
as $m\rightarrow\infty$.
Similarly, we can show that for any $b_{m}\rightarrow\infty$ and all $0\leq j\leq 4N$,
\begin{eqnarray}\label{a1-1}
\pr\Big{(}\sup_{0\leq t\leq G^{-1}(b_{m}/m^{o}_{2,\lambda_{j}})}\Big{|}\frac{\sum_{i\in\mathcal{H}_{0}}I\{|S_{i}|< \lambda_{j},|T_{i}|\geq t\}}{\hat{m}^{o}_{2,\lambda_{j}}G(t)}-1\Big{|}\geq \varepsilon\Big{)}\rightarrow 0
\end{eqnarray}
as $m\rightarrow\infty$.
By the definition of $\hat{t}_{1,\lambda}$, we have
\begin{eqnarray*}
\hat{m}_{1,\lambda}G(\hat{t}_{1,\lambda})=\alpha\max\Big{(}1,\sum_{i=1}^{m}I\{|S_{i}|\geq \lambda,|T_{i}|\geq \hat{t}_{1,\lambda}\}\Big{)}.
\end{eqnarray*}
Hence, by (\ref{a1-1-1}) and Lemma \ref{le5}, for any $\varepsilon>0$, $b_{m}\rightarrow\infty$,
\begin{eqnarray}\label{p3}
\pr\Big{(}FDP_{1,\lambda}(\hat{t}_{1,\lambda})\geq (1+\varepsilon)\alpha,m^{o}_{1,\lambda}G(\hat{t}_{1,\lambda})\geq b_{m}\Big{)}\rightarrow 0
\end{eqnarray}
as $m\rightarrow\infty$. Similarly, for any $0\leq j\leq 4N$, $\varepsilon>0$ and $b_{m}\rightarrow\infty$,
\begin{eqnarray}\label{p4}
\pr\Big{(}FDP_{2,\lambda_{j}}(\hat{t}_{2,\lambda_{j}})\geq (1+\varepsilon)\alpha,m^{o}_{2,\lambda_{j}}G(\hat{t}_{2,\lambda_{j}})\geq b_{m}\Big{)}\rightarrow 0
\end{eqnarray}
as $m\rightarrow\infty$. Define
\begin{eqnarray*}
FDP_{\lambda}=\frac{|\mathcal{R}_{0,\lambda}|}{\max(1,|\mathcal{R}_{\lambda}|)}.
\end{eqnarray*}
Then
\begin{eqnarray}\label{p5}
\pr\Big{(}FDP_{\lambda}\geq (1+\varepsilon)\alpha,m^{o}_{i,\lambda}G(\hat{t}_{i,\lambda})\geq b_{m},i=1,2\Big{)}\rightarrow 0.
\end{eqnarray}
It follows that
\begin{eqnarray}\label{p6}
&&\pr\Big{(}FDP_{\hat{\lambda}}\geq (1+\varepsilon)\alpha,m^{o}_{i,\hat{\lambda}}G(\hat{t}_{i,\hat{\lambda}})\geq b_{m},i=1,2\Big{)}\cr
&&\leq\sum_{j=0}^{4N}\pr\Big{(}FDP_{\lambda_{j}}\geq (1+\varepsilon)\alpha,m^{o}_{i,\lambda_{j}}G(\hat{t}_{i,\lambda_{j}})\geq b_{m},i=1,2\Big{)}\cr
&&\rightarrow 0.
\end{eqnarray}
Take $b^{2}_{m}=o(c_{m}\wedge m)$.
For $0\leq\lambda\leq 4\sqrt{\log m}$, by Lemma \ref{le4} and Markov's inequality,
\begin{eqnarray*}
&&\pr\Big{(}\sum_{i\in\mathcal{H}_{0}}I\{|S_{i}|\geq\lambda,|T_{i}|\geq \hat{t}_{1,\lambda}\}\geq b^{2}_{m},
m^{o}_{1,\lambda}G(\hat{t}_{1,\lambda})< b_{m}\Big{)}\cr
&&\leq\pr\Big{(}\sum_{i\in\mathcal{H}_{0}}I\{|S_{i}|\geq\lambda,|T_{i}|\geq G^{-1}(\min(1,b_{m}/m^{o}_{1,\lambda}))\}\geq b^{2}_{m}\Big{)}\cr
&&\leq C/b_{m}\cr
&&\rightarrow 0.
\end{eqnarray*}
Hence, we have
\begin{eqnarray*}
\pr\Big{(}\sum_{i\in\mathcal{H}_{0}}I\{|S_{i}|\geq\hat{\lambda},|T_{i}|\geq \hat{t}_{1,\hat{\lambda}}\}\geq b^{2}_{m},
m^{o}_{1,\hat{\lambda}}G(\hat{t}_{1,\hat{\lambda}})< b_{m}\Big{)}\rightarrow 0
\end{eqnarray*}
as $m\rightarrow\infty$. Similarly,
\begin{eqnarray*}
\pr\Big{(}\sum_{i\in\mathcal{H}_{0}}I\{|S_{i}|<\hat{\lambda},|T_{i}|\geq \hat{t}_{2,\hat{\lambda}}\}\geq b^{2}_{m},
m^{o}_{2,\hat{\lambda}}G(\hat{t}_{2,\hat{\lambda}})< b_{m}\Big{)}\rightarrow 0.
\end{eqnarray*}
By $\pr(|\mathcal{R}_{\hat{\lambda}}|\geq c_{m})\rightarrow 1$, (\ref{p3}) and (\ref{p4}),  it follows that
\begin{eqnarray*}
\pr\Big{(}FDP_{\hat{\lambda}}\geq (1+\varepsilon)\alpha,m^{o}_{i,\hat{\lambda}}G(\hat{t}_{i,\hat{\lambda}})< b_{m}\Big{)}\rightarrow 0
\end{eqnarray*}
for $i=1,2$ and any $\varepsilon>0$. This, together with (\ref{p6}), proves that $\pr(FDP_{\hat{\lambda}}\leq (1+\varepsilon)\alpha)\rightarrow 1$ as $m\rightarrow\infty$.
 \qed

The proofs of  Lemmas \ref{le6}-\ref{le5} are given in the supplementary material Liu (2014).

\begin{lemma}\label{le6} We have for any $M>0$,
\begin{eqnarray}\label{c9}
\pr(|S_{i}|\geq \lambda)=(1+o(1))\pr(|N(0,1)+h_{i}|\geq \lambda)+O(m^{-M})
\end{eqnarray}
and
\begin{eqnarray}\label{c99}
\pr(|T_{i}|\geq \lambda)=(1+o(1))G(t)+O(m^{-M}),
\end{eqnarray}
uniformly in $0\leq \lambda\leq 4\sqrt{\log m}$ and $i\in\mathcal{H}_{0}$.
For $0\leq j\leq 4N$,
\begin{eqnarray}\label{c10}
\pr(|S_{i}|<\lambda_{j})=(1+o(1))\pr(|N(0,1)+h_{i}|< \lambda_{j})+O(m^{-M})
\end{eqnarray}
uniformly in $i\in\mathcal{H}_{0}$.
\end{lemma}

\begin{lemma}\label{le4} We have for any $M>0$,
\begin{eqnarray}\label{lee4}
\pr(|S_{i}|\geq \lambda,|T_{i}|\geq t)=(1+o(1))\pr(|N(0,1)+h_{i}|\geq \lambda)G(t)+O(m^{-M})
\end{eqnarray}
uniformly in $0\leq \lambda\leq 4\sqrt{\log m}$, $0\leq t\leq 4\sqrt{\log m}$ and $i\in\mathcal{H}_{0}$. For all $0\leq j\leq 4N$,
\begin{eqnarray}\label{lee5}
\pr(|S_{i}|< \lambda_{j},|T_{i}|\geq t)=(1+o(1))\pr(|N(0,1)+h_{i}|< \lambda_{j})G(t)+O(m^{-M})
\end{eqnarray}
uniformly in $0\leq t\leq 4\sqrt{\log m}$ and $i\in\mathcal{H}_{0}$.
\end{lemma}

\begin{lemma}\label{le5} Let $b_{m}\rightarrow\infty$ be a sequence of positive numbers. (i). Assume that $\lambda$  satisfies $0\leq\lambda\leq 4\sqrt{\log m}$. We have
\begin{eqnarray}\label{le5-5}
\pr\Big{(}\Big{|}\frac{\hat{m}^{o}_{1,\lambda}}{m^{o}_{1,\lambda}}-1\Big{|}\geq\varepsilon\Big{)}I\{m^{o}_{1,\lambda}\geq b_{m}\}\rightarrow 0\quad\mbox{as $m\rightarrow\infty$.}
\end{eqnarray}
(ii). For $0\leq j\leq 4N$,
\begin{eqnarray}\label{le5-55}
\pr\Big{(}\Big{|}\frac{\hat{m}^{o}_{2,\lambda_{j}}}{m^{o}_{2,\lambda_{j}}}-1\Big{|}\geq\varepsilon\Big{)}I\{m^{o}_{2,\lambda_{j}}\geq b_{m}\}\rightarrow 0\quad\mbox{as $m\rightarrow\infty$.}
\end{eqnarray}
\end{lemma}

\subsection{Proof of Theorem \ref{th5}}

 By (\ref{a1}) with $\lambda=0$, we have for any $b_{m}\rightarrow\infty$ and $\varepsilon>0$,
\begin{eqnarray}\label{c9-9}
\pr\Big{(}\sup_{0\leq t\leq G^{-1}(b_{m}/m_{0})}\Big{|}\frac{\sum_{i\in\mathcal{H}_{0}}I\{|T_{i}|\geq t\}}{m_{0}G(t)}-1\Big{|}\geq \varepsilon\Big{)}\rightarrow 0.
\end{eqnarray}
The B-H method is equivalent to reject $H_{i0}$ if and only if $p_{i}\leq \hat{t}_{BH}$, where
\begin{eqnarray*}
\hat{t}_{BH}=\max\Big{\{}0\leq t\leq 1:~ mt\leq\alpha\max\Big{(}\sum_{i=1}^{m}I\{p_{i}\leq t\},1\Big{)}\Big{\}}
\end{eqnarray*}
and $p_{i}=G(|T_{i}|)$. By the definition of $\hat{t}_{BH}$, we have
\begin{eqnarray}\label{a19}
m\hat{t}_{BH}=\alpha\max(\sum_{i=1}^{m}I\{p_{i}\leq \hat{t}_{BH}\},1).
\end{eqnarray}
Let $\hat{R}_{BH}=\sum_{i=1}^{m}I\{p_{i}\leq \hat{t}_{BH}\}$ and $\hat{R}_{1,BH}=\sum_{i\in\mathcal{H}_{1}}I\{p_{i}\leq \hat{t}_{BH}\}$.
By (\ref{c9-9}) and $m_{1}=o(m)$, for any $\varepsilon>0$ and $c_{m}\rightarrow\infty$ with $b_{m}^{2}=o(c_{m})$ and $c_{m}=o(m_{1})$,
\begin{eqnarray*}
\pr\Big{(}\Big{|}\frac{FDP_{BH}}{\alpha}-1\Big{|}\geq \varepsilon, \hat{R}_{BH}\geq c_{m}\Big{)}\rightarrow 0\quad\mbox{as $m\rightarrow\infty$.}
\end{eqnarray*}
Hence, for any $\varepsilon_{1}>0$, we have
\begin{eqnarray}\label{c99}
\pr\Big{(}\hat{R}_{1,BH}\geq (1-\alpha+\varepsilon_{1}) \hat{R}_{BH},\hat{R}_{BH}\geq c_{m}\Big{)}\rightarrow 0\quad\mbox{as $m\rightarrow\infty$.}
\end{eqnarray}
Take $\varepsilon_{1}>0$ and $\varepsilon_{2}>0$ such that $(1-\varepsilon)(1-\alpha+\varepsilon_{1})\leq (1-\alpha-\alpha\varepsilon_{2})$.
Since $|\mathcal{R}_{\hat{\lambda}}|\geq \hat{R}_{BH}$, by the proof of Theorem 3.1, we have
$
\pr(FDP\geq (1+\varepsilon_{2})\alpha, \hat{R}_{BH}\geq c_{m})\rightarrow 0.
$
This implies that
\begin{eqnarray}\label{c109}
\pr(|\mathcal{R}_{\hat{\lambda}}|-|\mathcal{R}_{0,\hat{\lambda}}|\leq (1-\alpha-\alpha\varepsilon_{2})|\mathcal{R}_{\hat{\lambda}}|, \hat{R}_{BH}\geq c_{m})\rightarrow 0.
\end{eqnarray}
It follows from (\ref{c99}) and (\ref{c109}) that
\begin{eqnarray*}
\pr(power_{US}\leq (1-\varepsilon)power_{BH}, \hat{R}_{BH}\geq c_{m})\rightarrow 0
\end{eqnarray*}
as $m\rightarrow\infty$.
Note that $\pr(power_{BH}\geq \varepsilon,\hat{R}_{BH}\leq c_{m})\rightarrow 0$. So we have
$\pr(power_{US}\geq power_{BH}-\varepsilon)\rightarrow 1$ for any $\varepsilon>0$. Theorem 3.2 is proved. \qed

\subsection{Proof of Theorem \ref{th3}}

For $\varepsilon>0$, define
\begin{eqnarray*}
\F_{1}=\Big{\{}\Big{|}\frac{\sum_{i\in\mathcal{H}_{0}} I\{p_{i}\leq \hat{t}_{BH}\}}{m_{0}\hat{t}_{BH}}-1\Big{|}\leq \varepsilon\Big{\}}
\end{eqnarray*}
and $\E_{1}=\{\hat{t}_{BH}\geq m^{-w}\}$ for some $0<w<1$, where $m_{0}=|\mathcal{H}_{0}|$.  By (\ref{a1}), we have for any $0<w<1$,
\begin{eqnarray}\label{a25}
\sup_{m^{-w}\leq t\leq 1}\Big{|}\frac{\sum_{i\in\mathcal{H}_{0}} I\{p_{i}\leq t\}}{m_{0}t}-1\Big{|}\rightarrow 0
\end{eqnarray}
in probability as $m\rightarrow\infty$. Thus, $\pr(\E_{1}\cap\F^{c}_{1})=o(1)$. On $\F_{1}$, we have
\begin{eqnarray*}
m\hat{t}_{BH}=(\alpha+O(\varepsilon))\Big{(}\sum_{i\in\mathcal{H}_{1}}I\{p_{i}\leq\hat{t}_{BH}\}+m_{0}\hat{t}_{BH}\Big{)}
\end{eqnarray*}
which implies that
\begin{eqnarray}\label{a14}
\frac{\sum_{i\in\mathcal{H}_{1}}I\{p_{i}\leq\hat{t}_{BH}\}}{m\hat{t}_{BH}}=\alpha^{-1}-1+O(\varepsilon).
\end{eqnarray}
Hence, on $\F_{1}$, we have $\hat{t}_{BH}\leq C_{\alpha,\varepsilon}m^{\beta-1}$.
Let
\begin{eqnarray*}
T^{'}_{i}=\frac{\bar{X}_{i,1}-\bar{X}_{i,2}-(\mu_{i,1}-\mu_{i,1})}{\sqrt{\hat{\sigma}^{2}_{i,1}/n_{1}+\hat{\sigma}^{2}_{i,2}/n_{2}}}.
\end{eqnarray*}
For $i\in\mathcal{H}_{1}$, we have
\begin{eqnarray*}
I\{p_{i}\leq \hat{t}_{BH}\}&\leq& I\{T^{'}_{i}\geq G^{-1}(\hat{t}_{BH})-\frac{\sqrt{\sigma^{2}_{i,1}/n_{1}+\sigma^{2}_{i,2}/n_{2}}}{\sqrt{\hat{\sigma}^{2}_{i,1}/n_{1}+\hat{\sigma}^{2}_{i,2}/n_{2}}}\theta\sqrt{\log m}\}\cr
& &+I\{-T^{'}_{i}\geq G^{-1}(\hat{t}_{BH})-\frac{\sqrt{\sigma^{2}_{i,1}/n_{1}+\sigma^{2}_{i,2}/n_{2}}}{\sqrt{\hat{\sigma}^{2}_{i,1}/n_{1}+\hat{\sigma}^{2}_{i,2}/n_{2}}}\theta\sqrt{\log m}\}\cr
&=:&I_{i,1}+I_{i,2}.
\end{eqnarray*}
Since $\theta<\sqrt{2(1-\beta)}$, by central limit theorem and (2) in the supplementary material Liu (2014),
\begin{eqnarray*}
g_{m,i}:=\pr\Big{(}T^{'}_{i}\geq G^{-1}(C_{\alpha,\varepsilon}m^{\beta-1})
-\frac{\sqrt{\sigma^{2}_{i,1}/n_{1}+\sigma^{2}_{i,2}/n_{2}}}{\sqrt{\hat{\sigma}^{2}_{i,1}/n_{1}+\hat{\sigma}^{2}_{i,2}/n_{2}}}\theta\sqrt{\log m}\Big{)}\rightarrow 0
\end{eqnarray*}
uniformly in $i\in\mathcal{H}_{1}$.
By Markov's inequality,
\begin{eqnarray}\label{a13}
\pr\Big{(}\frac{\sum_{i\in\mathcal{H}_{1}}I_{i,1}}{m_{1}}\geq \varepsilon,\F_{1}\Big{)}
\leq\varepsilon^{-1}\frac{\sum_{i\in\mathcal{H}_{1}}g_{m,i}}{m_{1}}=o(1).
\end{eqnarray}
Similarly,
\begin{eqnarray*}
\pr\Big{(}\frac{\sum_{i\in\mathcal{H}_{1}}I_{i,2}}{m_{1}}\geq \varepsilon,\F_{1}\Big{)}=o(1).
\end{eqnarray*}
On $\E_{1}^{c}$, we have $\hat{t}_{BH}<m^{-w}\leq m^{\beta-1}$, where we take $1-\beta<w<1$. Hence, as in (\ref{a13}),
\begin{eqnarray*}
\pr\Big{(}\frac{\sum_{i\in\mathcal{H}_{1}}(I_{i,1}+I_{i,2})}{m_{1}}\geq \varepsilon,\E^{c}_{1}\Big{)}=o(1).
\end{eqnarray*}
This proves $power_{BH}\rightarrow 0$ in probability.

We next prove the theorem when $\theta>\sqrt{2(1-\beta)}$. So there exists some $\epsilon>0$ such that $\theta>\sqrt{2(1-\beta+\epsilon)}$.
Suppose that $\theta<\sqrt{2(1+\epsilon)}$.
By (\ref{a25}), we have
\begin{eqnarray*}
\frac{\sum_{i\in\mathcal{H}_{0}}I\{p_{i}\leq m^{-\theta^{2}/2+\epsilon}\}}{m_{0}m^{-\theta^{2}/2+\epsilon}}\rightarrow 1
\end{eqnarray*}
in probability as $m\rightarrow\infty$. Also, by by central limit theorem and (2) in the supplementary material Liu (2014), $\pr(|T_{i}|\geq G^{-1}(m^{-\theta^{2}/2+\epsilon}))\rightarrow 1$ uniformly for $i\in\mathcal{H}_{1}$. Hence
\begin{eqnarray*}
\frac{\sum_{i\in\mathcal{H}_{1}}I\{p_{i}\leq  m^{-\theta^{2}/2+\epsilon}\}}{m_{1}}\rightarrow 1
\end{eqnarray*}
in probability as $m\rightarrow\infty$. By $\theta>\sqrt{2(1-\beta+\epsilon)}$, we have $m^{1-\theta^{2}/2+\epsilon}=o(m_{1})$, and hence $\hat{t}_{BH}\geq m^{-\theta^{2}/2+\epsilon}$ with probability tending to one. This implies that
\begin{eqnarray*}
\frac{\sum_{i\in\mathcal{H}_{1}}I\{p_{i}\leq  \hat{t}_{BH}\}}{m_{1}}\rightarrow 1
\end{eqnarray*}
in probability as $m\rightarrow\infty$.
Suppose that $\theta\geq \sqrt{2(1+\epsilon)}$. Then we have $\pr(|T_{i}|\geq \sqrt{2\log m})\rightarrow 1$ uniformly for $i\in\mathcal{H}_{1}$.
This yields that
\begin{eqnarray*}
\frac{\sum_{i\in\mathcal{H}_{1}}I\{p_{i}\leq  G(\sqrt{2\log m})\}}{m_{1}}\rightarrow 1
\end{eqnarray*}
in probability as $m\rightarrow\infty$. By the definition of $\hat{t}_{BH}$, we have $\hat{t}_{BH}\geq \alpha/m\geq G(\sqrt{2\log m})$ when $m$ is large, which implies that
\begin{eqnarray*}
\frac{\sum_{i\in\mathcal{H}_{1}}I\{p_{i}\leq  \hat{t}_{BH}\}}{m_{1}}\rightarrow 1
\end{eqnarray*}
in probability as $m\rightarrow\infty$. The proof of the theorem is complete. \qed

\subsection{Proof of Theorem \ref{th4}}

We only need to prove the theorem when $\gamma<1$.
Let $\tau_{m}$ satisfy
\begin{eqnarray*}
\sum_{i\in\mathcal{H}_{0}^{'}}\pr(|N(0,1)+h_{i}|\geq \tau_{m}\sqrt{\log m})=m^{\beta^{*}},
\end{eqnarray*}
where $\beta^{*}=\beta+\min(1-\beta,\theta^{2}/4)$ and
\begin{eqnarray*}
\mathcal{H}_{0}^{'}=\Big{\{}i\in\mathcal{H}_{0}:~\sqrt{\frac{n_{1}}{\sigma^{2}_{i,1}(1+\frac{n_{2}\sigma^{2}_{i,1}}{n_{1}\sigma^{2}_{i,2}})}}
\Big{|}\mu_{i,1}+\frac{n_{2}\sigma^{2}_{i,1}}{n_{1}\sigma^{2}_{i,2}}\mu_{i,2}\Big{|}< h\sqrt{\log m}\Big{\}}.
\end{eqnarray*}
We have $\tau_{m}\leq h+\sqrt{2(1-\beta^{*})}+\varepsilon$ for any $\varepsilon>0$ when $m$ is large. Since $h\leq 2$,  there exists an $k^{*}$ such that $k^{*}/N\leq \tau_{m}\leq (k^{*}+1)/N$. Set
\begin{eqnarray*}
\sum_{i\in\mathcal{H}_{0}^{'}}\pr(|N(0,1)+h_{i}|\geq (k^{*}/N)\sqrt{\log m})= m^{q_{m}},
\end{eqnarray*}
where, by the tail probability of normal distribution, $q_{m}$ satisfies $|q_{m}-\beta^{*}|< (\tau_{m}+h)/N+1/N^{2}\leq 7/N$. Since $N\geq 10/\min(1-\beta,\theta^{2}/4)$, we have $
 q_{m}\geq \beta+\epsilon$ for some $\epsilon>0$.
By Lemma \ref{le6} and the proof of Lemma \ref{le5}, we can show that
\begin{eqnarray*}
\frac{\sum_{i\in\mathcal{H}_{0}^{'}}I\{|S_{i}|\geq (k^{*}/N)\sqrt{\log m}\}}{m^{q_{m}}}\rightarrow 1
\end{eqnarray*}
in probability. Hence $\pr(\hat{m}^{o}_{1,\lambda^{*}}\geq m^{q_{m}}/2)\rightarrow 1$ with $\lambda=(k^{*}/N)\sqrt{\log m}$ and $\hat{m}_{1,\lambda^{*}}/\hat{m}^{o}_{1,\lambda^{*}}\rightarrow 1$ in probability.
Put
\begin{eqnarray*}
\mathcal{H}^{'}_{1}=\Big{\{}i\in\mathcal{H}_{1}:~\sqrt{\frac{n_{1}}{\sigma^{2}_{i,1}(1+\frac{n_{2}\sigma^{2}_{i,1}}{n_{1}\sigma^{2}_{i,2}})}}
\Big{|}\mu_{i,1}+\frac{n_{2}\sigma^{2}_{i,1}}{n_{1}\sigma^{2}_{i,2}}\mu_{i,2}\Big{|}\geq \kappa\sqrt{\log m}\Big{\}}.
\end{eqnarray*}
We have $\kappa>\tau_{m}+\epsilon\geq k^{*}/N+\epsilon$ for some $\epsilon>0$.
Let $x_{m}=(\theta-\epsilon_{1})\sqrt{\log m}$ for some $\epsilon_{1}>0$ such that $\theta-\epsilon_{1}>\sqrt{\max(0,2\gamma-2\beta)}$.
It is easy to show that
\begin{eqnarray*}
\pr(|S_{i}|\geq (k^{*}/N)\sqrt{\log m})\rightarrow 1\quad\mbox{and\quad} \pr(|T_{i}|\geq x_{m})\rightarrow 1
\end{eqnarray*}
uniformly in $i\in\mathcal{H}_{1}^{'}$.
Hence
\begin{eqnarray*}
\pr(|S_{i}|\geq (k^{*}/N)\sqrt{\log m},|T_{i}|\geq x_{m})\rightarrow 1
\end{eqnarray*}
uniformly in $i\in\mathcal{H}_{1}^{'}$.
By Markov's inequality,
\begin{eqnarray*}
\frac{\sum_{i\in\mathcal{H}^{'}_{1}}I\{|S_{i}|\geq \lambda^{*},|T_{i}|\geq x_{m}\}}{|\mathcal{H}^{'}_{1}|}\rightarrow 1
\end{eqnarray*}
in probability. Since $N\geq 10/\min(1-\beta,\theta^{2}/4)$, we have $q_{m}\leq \beta^{*}+7/N\leq \beta+1.7\min(1-\beta,\theta^{2}/4)$.
Also, $\pr(\hat{m}_{1,\lambda^{*}}\leq m^{\gamma}+2m^{q_{m}})\rightarrow 1$.
By taking $\epsilon_{1}$ in $x_{m}$ sufficiently small, we have $\hat{m}_{1,\lambda^{*}}G(x_{m})=o(|\mathcal{H}^{'}_{1}|)$ as $|\mathcal{H}^{'}_{1}|\geq \rho m^{\beta}$. Hence
$\pr(\hat{t}_{1,\lambda^{*}}\leq x_{m})\rightarrow 1$.
So $\pr(|\mathcal{R}_{11,\lambda^{*}}|\geq (1-\varepsilon)\rho m^{\beta})\rightarrow 1$ for any $\varepsilon>0$, where
\begin{eqnarray*}
\mathcal{R}_{11,\lambda^{*}}=\Big{\{}i\in\mathcal{H}_{1}:~I\{|S_{i}|\geq \lambda^{*},|T_{i}|\geq \hat{t}_{1,\lambda^{*}}\}=1\Big{\}}.
\end{eqnarray*}
By the definition of $\hat{t}_{1,\lambda}$,
\begin{eqnarray*}
\hat{m}_{1,\lambda^{*}}G(\hat{t}_{1,\lambda^{*}})=\alpha\max(1,\sum_{i=1}^{m}I\{|S_{i}|\geq \lambda^{*},|T_{i}|\geq \hat{t}_{1,\lambda^{*}}\}).
\end{eqnarray*}
By (\ref{a1}),
 $FDP_{1,\lambda^{*}}(\hat{t}_{1,\lambda^{*}}))\rightarrow \alpha$ in probability. So $\pr(|\mathcal{R}_{1,\lambda^{*}}|\geq \rho(1-\alpha+\varepsilon)^{-1} m^{\beta})\rightarrow 1$ for any $\varepsilon>0$, where
 \begin{eqnarray*}
\mathcal{R}_{1,\lambda^{*}}=\Big{\{}1\leq i\leq m:~I\{|S_{i}|\geq \lambda^{*},|T_{i}|\geq \hat{t}_{1,\lambda^{*}}\}=1\Big{\}}.
\end{eqnarray*}
Since $|\mathcal{R}_{\hat{\lambda}}|\geq|\mathcal{R}_{\lambda^{*}}|$, it follows that
\begin{eqnarray*}
\pr\Big{(}|\mathcal{R}_{\hat{\lambda}}|\geq \rho(1-\alpha+\varepsilon)^{-1} m^{\beta})\rightarrow 1.
\end{eqnarray*}
By Theorem \ref{th1}, $\pr(FDP\leq \alpha+\varepsilon)\rightarrow 1$ for any $\varepsilon>0$. This implies that $\pr(power_{US}\geq \rho-\varepsilon)\rightarrow 1$ for any $\varepsilon>0$.\qed

\end{document}